\documentclass[
 aip,
 amsmath,amssymb,
preprint
]{./revtex4-2}
\usepackage{hyperref}
\usepackage{natbib}
\usepackage{graphicx}
\usepackage{dcolumn}
\usepackage{bm}
\usepackage{multirow}
\usepackage{amsmath}
\usepackage{makecell}
\usepackage{mathrsfs}
\usepackage{tikz}

\begin{document}

\preprint{ArXiv/manuscript}
\pagestyle{plain}

\title{Self-Supervised Learning for Effective Denoising of Flow Fields}

\author{Linqi Yu}
\affiliation{School of Mechanical Engineering, Pusan National University, 2, Busandaehak-ro 63beon-gil, Geumjeong-gu, Busan, 46241, Rep. of KOREA}

\author{Mustafa Z. Yousif}
\affiliation{School of Mechanical Engineering, Pusan National University, 2, Busandaehak-ro 63beon-gil, Geumjeong-gu, Busan, 46241, Rep. of KOREA}

\author{Dan Zhou}
\affiliation{School of Mechanical Engineering, Pusan National University, 2, Busandaehak-ro 63beon-gil, Geumjeong-gu, Busan, 46241, Rep. of KOREA}

\author{Meng Zhang}
\affiliation{School of Mechanical Engineering, Pusan National University, 2, Busandaehak-ro 63beon-gil, Geumjeong-gu, Busan, 46241, Rep. of KOREA}

\author{Jung Sub Lee}
\affiliation{Biomedical Research Institute, Department of Orthopaedic Surgery, Pusan National University Hospital, 179 Gudeok-Ro, Seo-gu, Busan, 46241, Rep. of KOREA}

\author{Hee-Chang Lim}
\email[]{Corresponding author, hclim@pusan.ac.kr}
\thanks{}
\affiliation{School of Mechanical Engineering, Pusan National University, 2, Busandaehak-ro 63beon-gil, Geumjeong-gu, Busan, 46241, Rep. of KOREA}
\email{hclim@pusan.ac.kr}

\date{\today}

\begin{abstract}
In this study, we proposed an efficient approach based on a deep learning (DL) denoising autoencoder (DAE) model for denoising noisy flow fields. The DAE operates on a self-learning principle and does not require clean data as training labels. Furthermore, investigations into the denoising mechanism of the DAE revealed that its bottleneck structure with a compact latent space enhances denoising efficacy. Meanwhile, we also developed a deep multiscale DAE for denoising turbulent flow fields. Furthermore, we used conventional noise filters to denoise the flow fields and performed a comparative analysis with the results from the DL method. The effectiveness of the proposed DL models was evaluated using direct numerical simulation data of laminar flow around a square cylinder and turbulent channel flow data at various Reynolds numbers. For every case, synthetic noise was augmented in the data. A separate experiment used particle-image velocimetry data of laminar flow around a square cylinder containing real noise to test DAE denoising performance. Instantaneous contours and flow statistical results were used to verify the alignment between the denoised data and ground truth. The findings confirmed that the proposed method could effectively denoise noisy flow data, including turbulent flow scenarios. Furthermore, the proposed method exhibited excellent generalization, efficiently denoising noise with various types and intensities.

\end{abstract}

\maketitle

\section{Introduction}
\label{Introduction} 
Understanding, modeling, and controlling fluid flow is crucial in various fields, such as energy production, aerospace engineering, and weather prediction. However, the fluid dynamics in these fields are complex. The turbulent flow is a particularly challenging problem because of its highly nonlinear and chaotic behavior. In most situations, mass and accurate data are required for fluid flow visualization and statistical analysis. Experimental and computational fluid dynamics (CFD) has been used to develop many efficient data generation methods. In CFD, direct numerical simulation (DNS) has been used to accurately simulate various cases of fluid flows by solving the Navier–Stokes equation\cite{Moin&Mahesh1998, Kimetal1987, Moseretal1999}. In experimental fields, particle-image velocimetry (PIV) was applied to visualize and quantify the complex instantaneous structure of fluid flows\cite{Adrian2005}. Advanced derivatives of PIV, such as tomographic PIV\cite{Scarano2013} and thermographic PIV\cite{Allison&Gillies1997}, are used to measure three-dimensional flow and temperature fields. However, the accuracy of PIV measurements is limited by white noise, such as mechanical vibration, inadequate illumination, background speckle, and optical problems caused by the inevitable errors of the experimental setup\cite{Scherletal2020}. Noise reduces the signal-to-noise ratio (SNR) of small-scale velocity fluctuations and PIV measurement accuracy, especially when measuring the small-scale velocity fluctuations of turbulent flows.\par

Several denoising methods, including energy, convolution, wavelet, and optimal Wiener filters, have been proposed to increase PIV measurement accuracy\cite{Vételetal2011, Oxladeetal2012}. In addition to filter-based denoising methods, data-driven approaches, such as proper orthogonal decomposition\cite{Lumley1967, Berkoozetal1993} and dynamic mode decomposition\cite{Schmid2010}, achieve excellent resolution reconstruction and noise elimination \cite{He&Liu2017, Scherletal2020}. Because of their linear nature, these methods exhibit a limited capacity for denoising the complex fluid flow with high nonlinearities and multiple spatiotemporal scales\cite{Bruntonetal2020}.\par

Deep learning (DL) algorithms, a subset of machine learning (ML), are increasingly being used in various domains\cite{Pouyanfaretal2018}. DL is widely adopted in fluid dynamics because it effectively handles highly nonlinear mappings\cite{Bruntonetal2020}. In particular, DL can be used to address challenges related to turbulent flow problems, such as the temporal data generation of turbulent flow\cite{Jiangetal2021, Duraisamyetal2019}, reconstruction of flow fields\cite{Yousifetal2021, Yousifetal2022d, Shaetal2023, Yuetal2022}, and prediction of flow field parameters\cite{Guastonietal2021, Yuetal2023}. \par

Using DL algorithms for data denoising is a critical topic of research. In computer vision studies, denoising DL models have been extensively investigated, especially for image denoising\cite{Tianetal2020}. However, most DL denoising models are based on supervised learning, which requires paired clean and noisy data to train DL models. Supervised DL methods are not always feasible in fluid flow dynamics problems because paired data are not always available. Therefore, unsupervised learning with unpaired datasets is an excellent alternative for reconstructing fluid flow fields. Generative adversarial networks (GANs) proposed by Goodfellow {\it et al.}\cite{Goodfellowetal2020}, is a well-known unsupervised learning algorithm. Based on the GANs, Kim {\it et al.}\cite{Kimetal2021} applied an unsupervised cycle-consistent GANs (CycleGANs) model for the super-resolution reconstruction of turbulent flows. However, although the low-resolution (LR) and high-resolution (HR) datasets are not paired in the training process, HR data are still required.\par 

Physics-informed neural networks (PINNs)\cite{Raissietal2019} have become a critical topic of interest. PINNs are function approximators described using partial differential equations for embedding the information of physical laws governing a given dataset in the learning process. Fathi {\it et al.}\cite{Fathietal2020} proposed a physics-informed DL method for processing 4D-flow MRI to increase the spatiotemporal resolution and reduce noise. This method used a deep neural network to approximate the underlying flow field. Similarly, Gao {\it et al.}\cite{Gaoetals2021} proposed a physics-informed deep learning (DL) solution for the spatial super-resolution of flow fields. DL model training only requires LR samples instead of their HR counterparts as labels using fluid flows' physical laws and boundary conditions. After adequate training, the DL model could spatially reconstruct the flow field. Here, a noisy LR was inputted in the parameter space. However, PINN limitations cannot be neglected. For example, PINNs cannot effectively propagate information from initial or boundary conditions to latent parts of the domain or future times, particularly in large computational domains, for example, unsteady turbulent flow\cite{Lietal2023, Faroughietal2022}. Therefore, PINNs can be weak in efficiently solving complex turbulent flow problems. \par 

Reinforcement learning (RL), another subset of the ML method in which an agent is trained to make decisions through interaction with the environment, achieves excellent results in robotics, game playing, natural language processing, and computer vision problems\cite{Li2018}. This learning approach renders deep RL (DRL) a viable method for addressing various challenges within fluid dynamics, such as active flow control \cite{Rabaultetal2019,Yousifetal2023a}, design optimization\cite{Viqueratetal2021}, and CFD\cite{Novatietal2021}. Yousif {\it et al.}\cite{Yousifetal2023b} proposed a physics-constrained DRL (PCDRL) to correct flow fields from noisy data. The momentum equation, pressure Poisson equation, and boundary conditions were used to calculate the reward function. Using the physics-based reward function, the agent can determine the most suitable filter to denoise the value for the points in the flow field. Using physical constraints, PCDRL was data-free without using clean data as the target.\par 

However, these physics-constrained and physics-informed methods have a common limitation: all flow field parameters are required to solve the governing equations. Therefore, they cannot solve certain cases, particularly unsteady turbulent flows in which two-dimensional (2D) data are derived from three-dimensional (3D) data, such as 2D experimental data obtained through PIV. This study developed an efficient method for fluid flow denoising without clean data requirements. This method is suitable for denoising noisy turbulent flow data.\par 

This study revealed that a bottleneck-shaped autoencoder neural network with a small latent space exhibits excellent flow field denoising ability. The autoencoder is a self-supervised DL model with the same input and target data. Thus, this method does not require clean data. This study investigated the mechanism by which the bottleneck-shaped autoencoder denoises data. To evaluate denoising performance, noisy data of laminar flow around a square cylinder were considered. Next, deep multiscale DAE (DMS-DAE) was developed to denoise the 2D sections of the noisy turbulent channel flow. Here, DNS generated the data of the flow cases, and noise was added synthetically. Noisy PIV data of flow over a square cylinder were applied to check the denoising ability of the proposed DAE model on real experimental data.\par 

The remainder of this paper is organized as follows. Sec. 2 introduces DNS data generation, the methods for adding synthetic noise, and the experimental data generation used in this study. Sec. 3 details the proposed model. Sec. 4 details the results for all cases. Finally, Sec. 5 presents the conclusions.\par

\section{Generation of the data set}
\label{Data Generation}
\subsection{DNS data and synthetic noise}
\label{DNS data and synthetic noise}

This study used two DNS flow data sets with synthetic noise to evaluate the DL model. First, the 2D laminar flow around a square cylinder (bluff body flow) at $Re_D$ = 100 were used as a demonstration case. The turbulent channel flow at $Re_\tau$ = 180, 395, and 550 were used for subsequent denoising performance tests of MS-DAE. \par

In the bluff body flow, the simulation domain size was set to $x_D$ × $y_D$  = 20$D$ × 15$D$, where $x$ and $y$ are the streamwise and spanwise directions, respectively. The corresponding grid size was 381 × 221. The stretching mesh technique applied local mesh refinement near the cylinder walls. The uniform inlet velocity and pressure outlet boundary conditions were applied to the inlet and outlet of the domain. Furthermore, no-slip boundary conditions were used in the cylinder walls and the symmetry plane to the sides of the domain. The dimensionless time step of the simulation, $u_\infty \Delta t/D$, was set to $10^2$. The DNS data used in this study were cut around the edges and interpolated into uniform grids, where the new domain size is $x_D$ × $y_D$  = 17$D$ × 8$D$ and the uniform grid size is 256 × 128. Moreover, the interval between the collected snapshots of the flow fields i s ten times the duration of the simulation time step. A total of 1200 snapshots were collected and used.\par

For the turbulent channel flow, three friction Reynolds numbers, namely $Re_\tau$ = $u_\tau\delta/\nu$ = 180, 395, and 550, were used. Here, $u_\tau$ is the friction velocity, and $\delta$ is half of the channel height. Table 1\ref{tab:Table1} lists the simulation parameters of each $Re_\tau$. The periodic boundary condition was used in the streamwise and spanwise directions. Moreover, the no-slip condition was applied to the upper and lower walls of the channel. The turbulence statistics obtained from the DNS were validated by comparing them with the results from reference paper\cite{Kimetal1987, Moseretal1999}.

\begin{table}
  \begin{center}
\scalebox{0.7}{\renewcommand\arraystretch{1.5}
\begin{tabular}{cccccccccc} \hline\hline
&$Re_\tau$~ ~&~~$L_x\times L_y \times L_z$~ ~&~~$N_x\times N_y \times N_z$~ ~&~~$\Delta x^+$~ ~&~~$\Delta z^+$~ ~&~~$\Delta y_w^+$~ ~&~~$\Delta y_c^+$~ ~&~~$\Delta t^+$&  \\ \hline
&$180$~ ~&~~$4\pi\delta\times2\delta\times2\pi\delta$~~&~~$256 \times 128 \times 256$~~&~~$8.831$~~&~~$4.415$~~&~~$0.63$~~&~~$4.68$~~&~~$0.113$&   \\
&$395$~ ~&~~$4\pi\delta\times2\delta\times2\pi\delta$~~&~~$385 \times 257 \times 385$~~&~~$12.553$~~&~~$6.277$~~&~~$0.541$~~&~~$5.115$~~&~~$0.023$& \\
&$550$~ ~&~~$4\pi\delta\times2\delta\times2\pi\delta$~~&~~$512 \times 336 \times 512$~~&~~$13.492$~~&~~$6.746$~~&~~$0.401$~~&~~$5.995$~~&~~$0.030$& \\ 
\hline\hline
\end{tabular}}
  \caption{DNS parameters of turbulent channel flows. Here, $L$ and $N$ are the domain dimension and grid number, respectively. The superscript + indicates that the quantity is nondimensionalized by $u_\tau$ and $\nu$. Here, $\Delta y_w^+$ and $\Delta y_c^+$ are the spacing near the wall and at the center of the channel, respectively.}
  \label{tab:Table1}
  \end{center}
\end{table}

Training and testing data were collected from the $y$–$z$ plane of the 3D channel flow domain. The flow data at $Re_\tau$ = 395 and 550 were interpolated to align with the grid size of the DNS data for the flow at $Re_\tau$ = 180, which features 128 and 256 grids in the $y$- and $z$-directions, respectively. This data processing ensures that the flow at all three Reynolds numbers can be input into the DL model simultaneously.\par

\begin{figure}
\centering 
\includegraphics[angle=0, trim=0 0 0 0, width=1\textwidth]{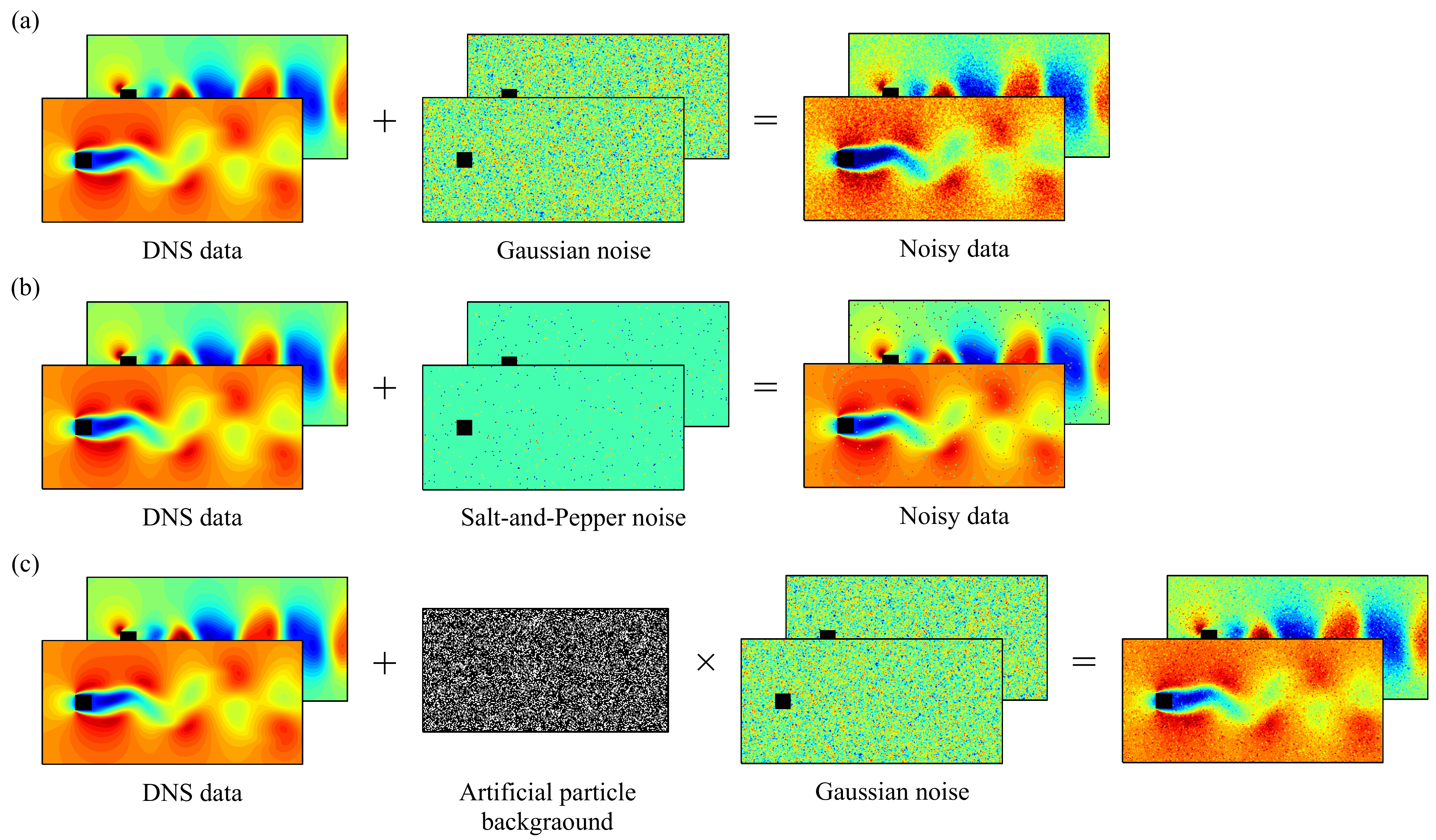}
\caption[]{Three types of synthetic noise: (a) Gaussian noise, (b) Salt-and-pepper noise, and (c) Speckle noise.}
\label{fig:1}
\end{figure}

In this study, synthetic noises, namely Gaussian noise, salt-and-pepper noise, and speckle noise, were added to degrade the quality of all DNS data cases. For Gaussian noise, zero-mean random noise with $\mathcal{N}$(0,$\sigma^2$) was added to the flow fields, as displayed in Figure \ref{fig:1} (a), where $\mathcal{N}$, $\sigma^2$ are the normal distribution and variance, respectively. Salt-and-pepper noise data are generated by randomly adding the minimum and maximum values of the velocity components to the flow data, as displayed in Figure \ref{fig:1} (b). Speckle noise is a granular texture degrading quality due to interference among wavefronts in coherent imaging systems, such as optical coherence tomography \cite{Lecompteetal2005}. This study simplified speckle noise by assuming it exists only at the particle positions in the PIV system. Thus, artificial speckle noise is generated using a method displayed in Figure \ref{fig:1} (c). First, an artificial particle background, where the density of the particles was 50\%, imitating the real PIV measurement, was devised. Next, all pixels with particles (white) were considered as coefficient 1. By contrast, the black region without particles was coefficient 0. Finally, speckle noise data were obtained by multiplying the background coefficient with Gaussian noise and adding it to the flow data. Moreover, the SNR represents the noise level. Here, SNR=$\sigma_{DNS}^2/\sigma_{noise}^2$, where $\sigma_{DNS}^2$ and $\sigma_{noise}^2$ are the variances of the DNS and noise data, respectively. This study defines the noise level as the reciprocal of the SNR, 1/SNR.\par

\subsection{Configuration of PIV measurement}
\label{Configuration of PIV measurement}

Noisy PIV data of flow over a square cylinder were generated and used in this study. Noisy PIV data were generated through a return-type water channel. The test section size of the water channel was 1 m (length) × 0.35 m (height) × 0.3 m (width). The free stream velocity was set to 0.02 m s-1 with an $Re_D$ of 200. Background noise was generated at high levels because of external noise and the sparse honeycomb configuration of the water channel. Polyamide12 seed particles of INTECH SYSTEMS were used to seed the channel with a 50 $\mu$m diameter. A high-speed camera (FASTCAM Mini UX 50) and a continuous laser with a 532 nm wavelength were used in the PIV system. The snapshot frequency was set to 24 Hz. In this experiment, a square cylinder model was constructed using an acrylic board, and the cross-section of the model was set to 1 × 1 cm. The model was not entirely transparent. Thus, a shadow region existed near the bluff body when the laser passed through the model.\par

\section{DL methodology}
\label{DL methodology}
\subsection{Denoising autoencoder}
\label{Denoising autoencoder}

Autoencoder (AE) is a classical neural network extensively used for image classification, object detection, and natural language processing \cite{Lietal2023}. AE solves various problems in fluid dynamics, such as the super-resolution reconstruction of flow fields \cite{Fukamietal2019b} and predicting turbulent dynamics \cite{Raccaetal2023}. Furthermore, in this study, AE's fluid flow denoising ability was introduced. The denoising mechanism of the autoencoder was investigated by comparing various networks with different structures. This comparison used 1000 and 200 snapshots of noisy bluff body flow data with Gaussian noise at 1/SNR = 0.5 as training and validation data, respectively. The training of the AE was based on self-supervised learning without clean label data, in which the input and output were the same. Thus, the testing data were the same as the training data. Table \ref{tab:Table2} summarises the comparison, which indicates that only bottleneck-shaped AE with convolutional neural networks (CNNs) and a small latent space size can denoise the data. Such AE is called DAE.\par

\begin{table}[]
\centering 
\scalebox{0.70}{\renewcommand\arraystretch{1.5}
\begin{tabular}{|ccc|c|c|}
\hline
\multicolumn{3}{|c|}{Cases}                                                                                                    & \multirow{2}{*}{Shape of networks} & \multirow{2}{*}{Denoising ability} \\ \cline{1-3}
\multicolumn{1}{|c|}{Network structure}                     & \multicolumn{1}{c|}{Type of layers}          & Latent sapce size &                                    &                                    \\ \hline
\multicolumn{1}{|c|}{\multirow{2}{*}{Non-AE}}               & \multicolumn{1}{c|}{Dense Layers}            & /                 & \multirow{2}{*}{\begin{minipage}[b]{0.3\textwidth}{\includegraphics[width=\linewidth]{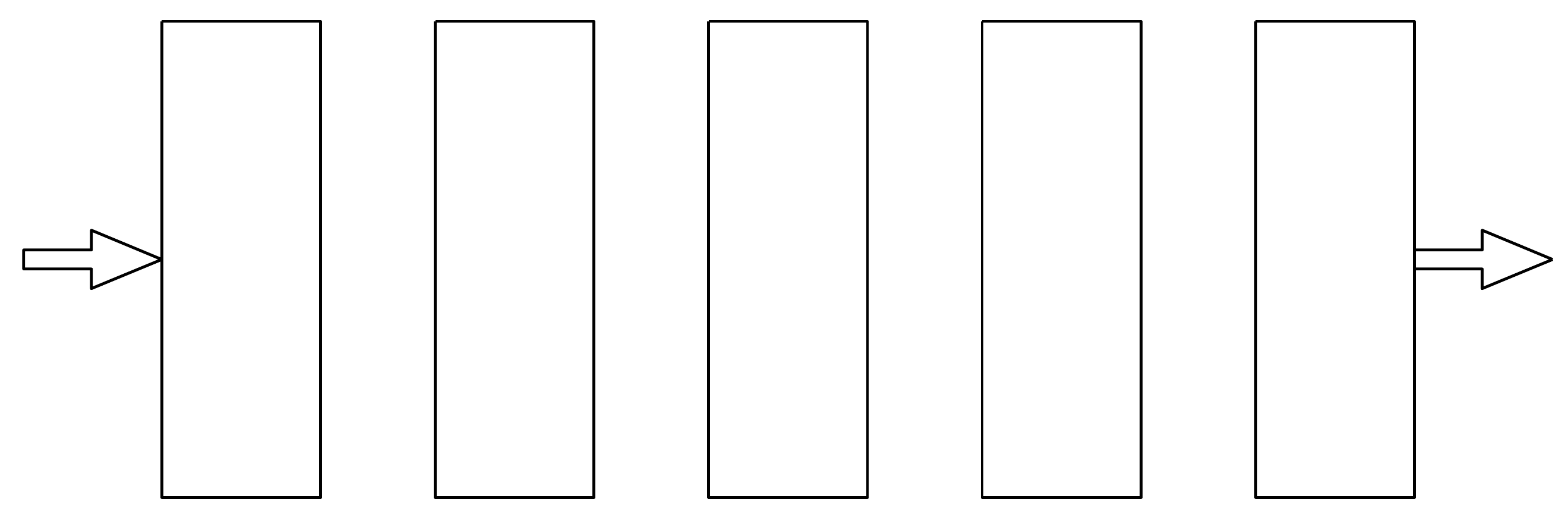}}   \end{minipage}   }                 & No                                 \\ \cline{2-3} \cline{5-5} 
\multicolumn{1}{|c|}{}                                      & \multicolumn{1}{c|}{CNN Layers}              & /                 &                                    & No                                 \\ \hline
\multicolumn{1}{|c|}{\multirow{3}{*}{Bottlebody-shaped AE}} & \multicolumn{1}{c|}{\multirow{3}{*}{CNN AE}} & 256 × 512         & \multirow{3}{*}{\begin{minipage}[b]{0.3\textwidth}{\includegraphics[width=\linewidth]{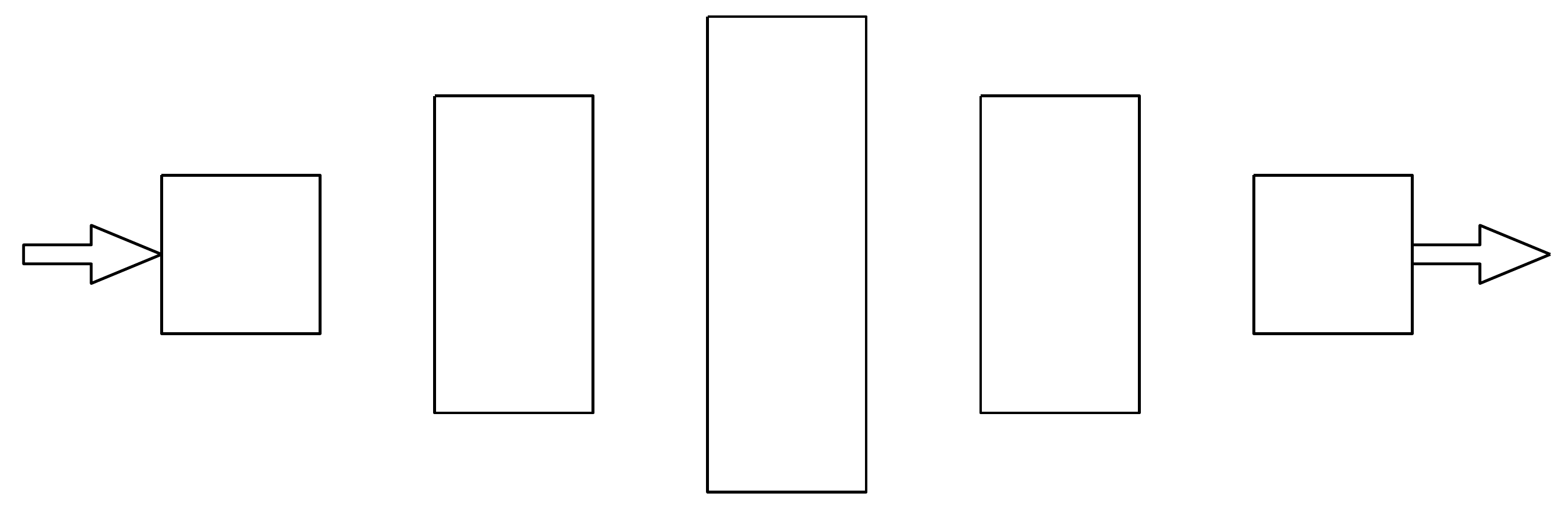}}   \end{minipage}   }           & No                                 \\ \cline{3-3} \cline{5-5} 
\multicolumn{1}{|c|}{}                                      & \multicolumn{1}{c|}{}                        & 512 × 1024        &                                    & No                                 \\ \cline{3-3} \cline{5-5} 
\multicolumn{1}{|c|}{}                                      & \multicolumn{1}{c|}{}                        & 1024 × 2048       &                                    & No                                 \\ \hline
\multicolumn{1}{|c|}{\multirow{6}{*}{Bottleneck-shaped AE}} & \multicolumn{1}{c|}{Dense Layers}            & 8 × 16            & \multirow{6}{*}{\begin{minipage}[b]{0.3\textwidth}{\includegraphics[width=\linewidth]{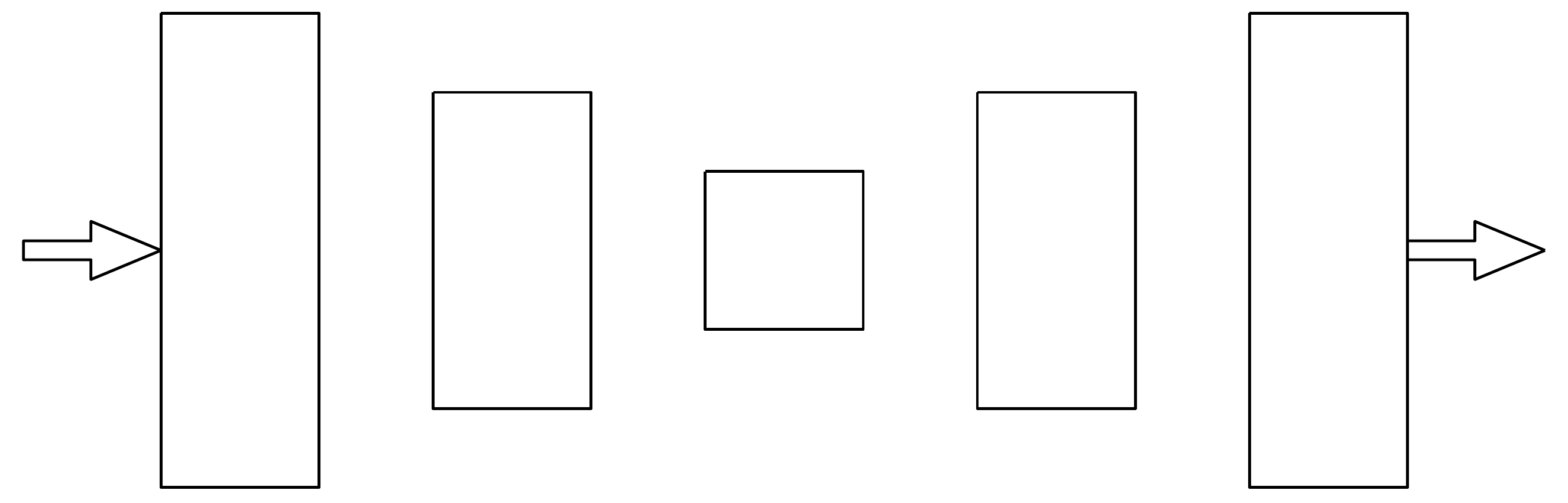}}   \end{minipage}   }                 & No                                 \\ \cline{2-3} \cline{5-5} 
\multicolumn{1}{|c|}{}                                      & \multicolumn{1}{c|}{\multirow{5}{*}{CNN AE}} & 64 × 128          &                                    & No                                 \\ \cline{3-3} \cline{5-5} 
\multicolumn{1}{|c|}{}                                      & \multicolumn{1}{c|}{}                        & 32 × 64           &                                    & No                                 \\ \cline{3-3} \cline{5-5} 
\multicolumn{1}{|c|}{}                                      & \multicolumn{1}{c|}{}                        & 16 × 32           &                                    & Yes                                \\ \cline{3-3} \cline{5-5} 
\multicolumn{1}{|c|}{}                                      & \multicolumn{1}{c|}{}                        & 8 × 16            &                                    & Yes                                \\ \cline{3-3} \cline{5-5} 
\multicolumn{1}{|c|}{}                                      & \multicolumn{1}{c|}{}                        & 4 × 8             &                                    & Yes                                \\ \hline
\end{tabular}}

  \caption{Influence of neural networks with various structures on flow field denoising performance}
  \label{tab:Table2}

\end{table}

\begin{figure}
\centering 
\includegraphics[angle=0, trim=0 0 0 0, width=0.8\textwidth]{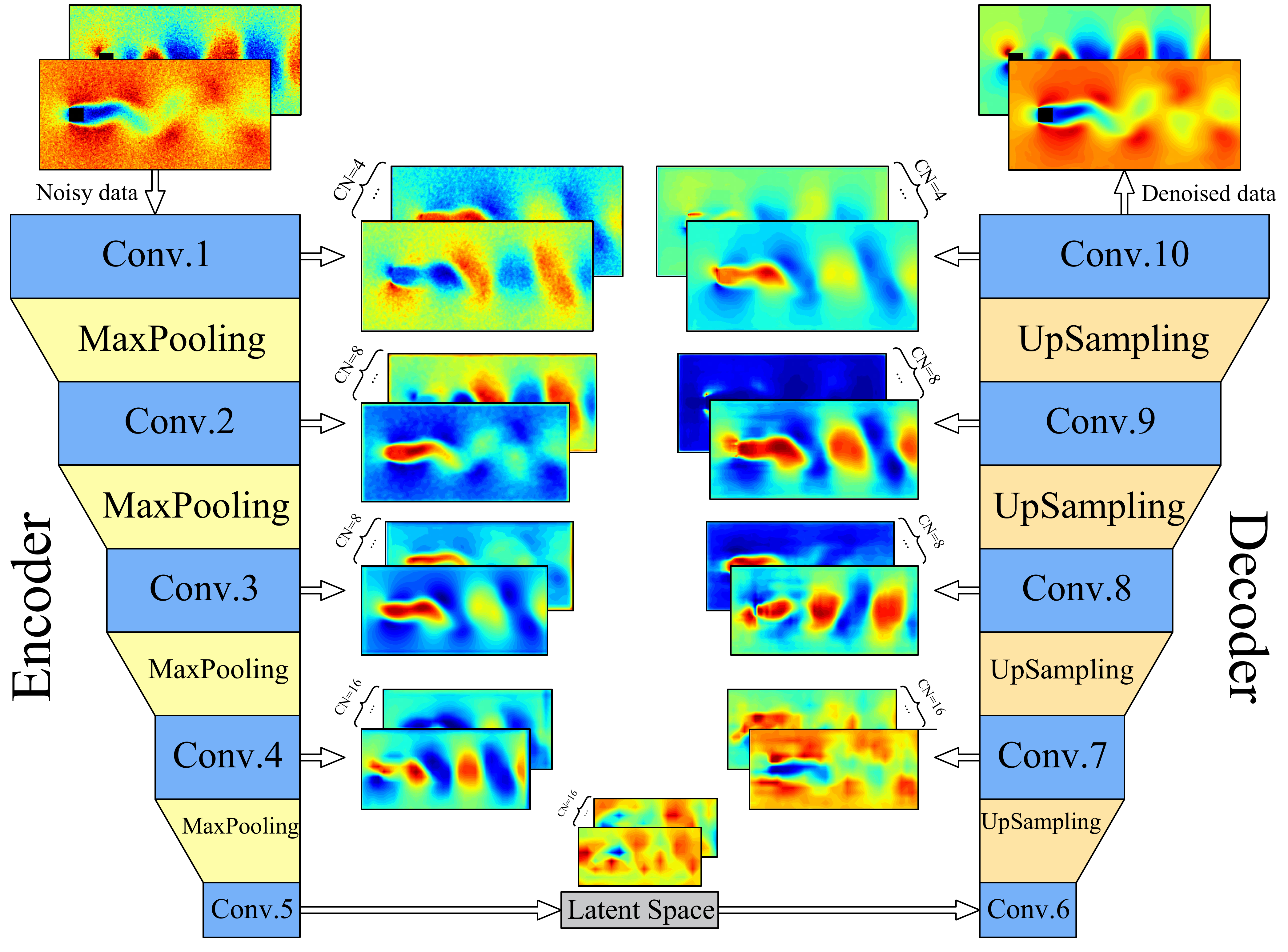}
\caption[]{DAE architecture.}
\label{fig:2}
\end{figure}

\begin{figure}
\centering 
\includegraphics[angle=0, trim=0 0 0 0, width=0.9\textwidth]{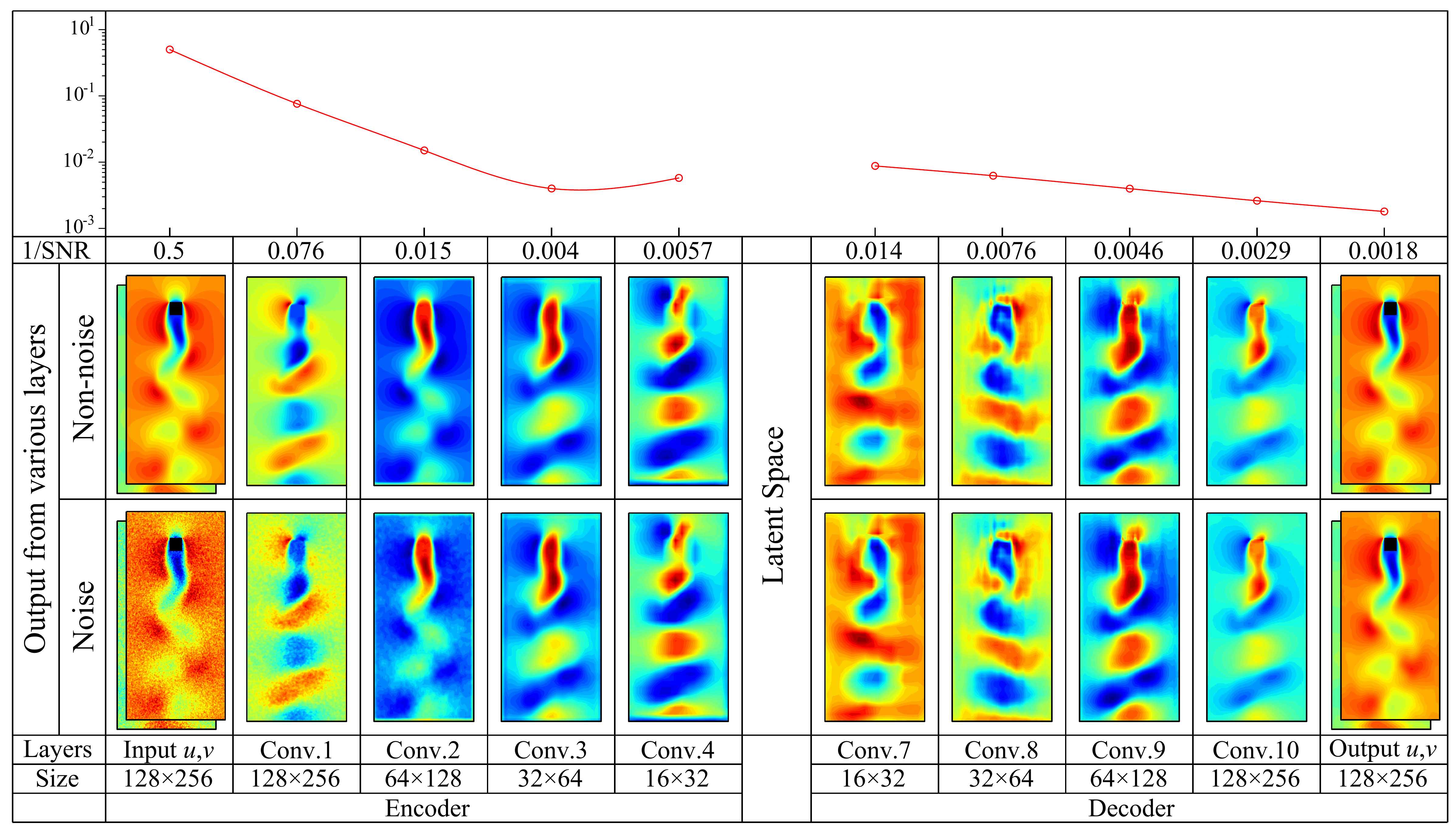}
\caption[]{Output analysis from various layers of the DAE.}
\label{fig:3}
\end{figure}

Figure \ref{fig:2} displays the structure of the DAE. The convolution and carpooling layers make up the encoder part of DAE. The decoder is composed of convolution and sampling layers. The channel number of the convolution layers varies with the AE depth, where the CH is larger when the convolution layers are closer to the latent space. Furthermore, noisy and clean data were passed several times into the DAE to output procedure data from various convolution layers. As displayed in Figure \ref{fig:3}, the noise level was reduced with the output from various convolution layers in the encoder. This phenomenon proved that the DAE encoder can filter the noise as a funnel. Therefore, only bottleneck-shaped AE can denoise rather than bottlebody-shaped AE. When noisy data passed the encoder, the data size was compressed and decreased by pooling layers. The filters of the convolution layer were not sufficiently sensitive for capturing high-frequency noise features. Thus, noise information cannot be preserved in the encoder. If noisy data were not compressed and filtered sufficiently with a large latent space, noise information was retained so that the AE loses its denoising ability. The noise of the output around the latent space increased slightly because the procedure data around the latent space were highly compressed and sensitive to 1/SNR calculation. The procedure data in the decoder exhibited a low noise level. Finally, the denoised data were reconstructed.\par

\subsection{Deep multiscale denoising autoencoder}
\label{Deep multiscale denoising autoencoder}

\begin{figure}
\centering 
\includegraphics[angle=0, trim=0 0 0 0, width=1\textwidth]{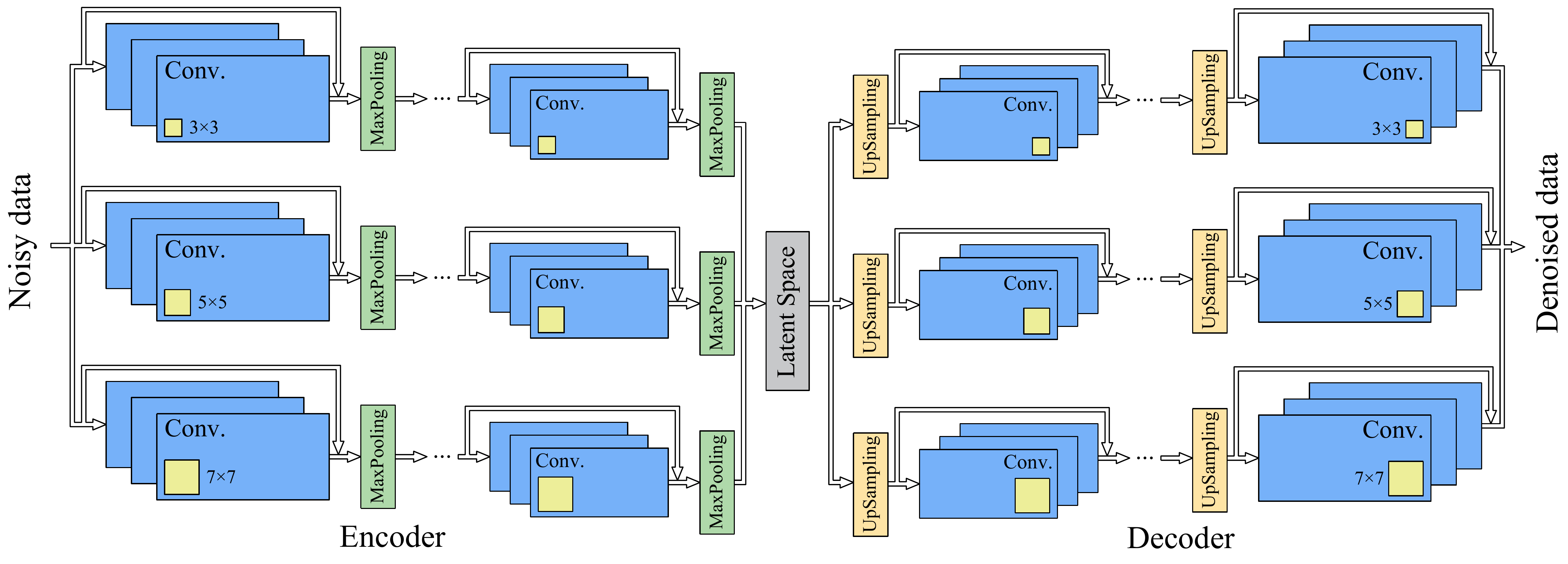}
\caption[]{DMDAE architecture.}
\label{fig:4}
\end{figure}

The DAE in the last subsection has a simple frame that can only denoise uncomplex flow fields. The performance of this DAE for denoising the turbulent flow was not satisfactory, as the DL model could not distinguish small eddies from noise and could not reconstruct the turbulent flow well. Thus, a complex deep multiscale DAE (DMDAE) was developed. As shown in Figure \ref{fig:4}, DMDAE features deep convolutional layers, enhancing its capability to capture features. The skip connexion was used to avoid the vanishing gradient problem in the training process. Furthermore, the encoder and decoder have three branches with various filter sizes. This multiscale CNN strategy renders the convolutional layers distinguishable \cite{Huangetal2019}.\par

This subsection introduces the training and testing processes of DAE and DMDAE. Before training, all data were normalized using the min–max normalization function using the minimum and maximum values from the average flow field to scale the values between 0 and 1. Considering DAE as an example, noisy data were passed into DAE to obtain output data. Next, the training loss was calculated from the output and noisy input data using the mean square error loss function. After calculating the training loss, the optimization algorithm updated the weights of the DAE to minimize the loss throughout the training period. This study used the adaptive moment estimation (Adam) algorithm as the optimization algorithm \cite{Kingma&Ba2017}. Validation data were used to validate the DL model by calculating the validation loss. The weights were saved only if the validation loss decreased compared with the previous epoch. On completion of the training process, testing data (identical to the training data) were inputted into the trained DL model to obtain denoised flow data. This study used the open-source library TensorFlow 2.3.0 to implement the DL model. The customized sample Python code for the proposed models can be accessed on the website (https://fluids.pusan.ac.kr/fluids/65416/subview.do).\par

\section{Results and discussion}
\label{Results and discussion} 

\begin{figure}
\centering 
\includegraphics[angle=0, trim=0 0 0 0, width=0.8\textwidth]{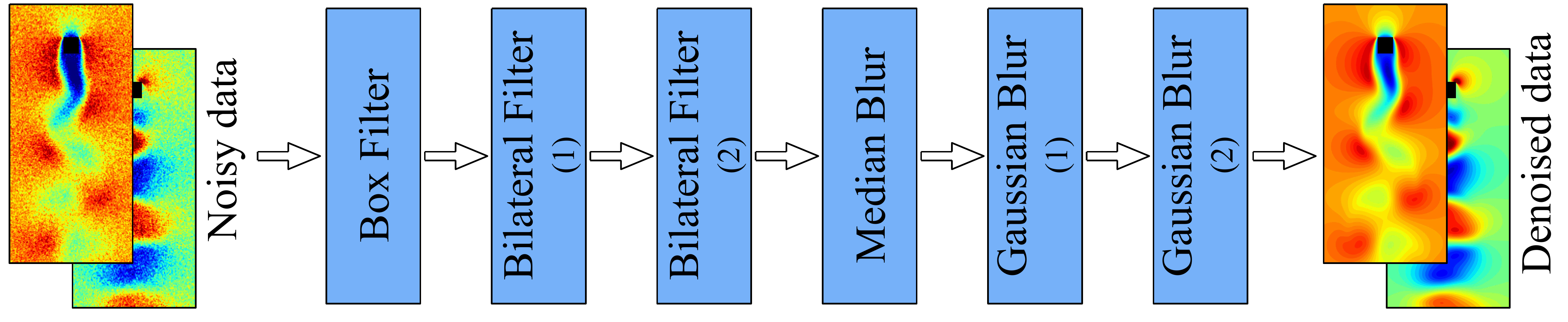}
\caption[]{Schematic of the CF.}
\label{fig:5}
\end{figure}

Four experiments were conducted to examine the denoising efficacy of the proposed self-supervised DL method. The first evaluation involved denoising bluff body flow data contaminated with Gaussian noise at various noise levels using DAE. The second evaluation focused on denoising bluff body flow data contaminated with three noise types, namely Gaussian, salt-and-pepper, and speckle, at a consistent noise level by using DAE. In the third evaluation, DMDAE eliminated noise from channel flows under noisy conditions at three Reynolds numbers. The fourth evaluation demonstrated denoising noisy PIV data. Furthermore, a comparative analysis was performed using a denoising method using classical filters (CF) and the proposed DL method. As displayed in Figure \ref{fig:5}, CF contains several denoising filters, including one box filter, two bilateral filters, one median blur filter, and two Gaussian blur filters. Table \ref{tab:Table3} plots the parameter settings of the CF.\par

\begin{table}
  \begin{center}
\scalebox{0.7}{\renewcommand\arraystretch{1.5}
\begin{tabular}{cccccccccc} \hline\hline
Classical filters & Parameters \\ \hline
Box filter & Filter size = 5 × 5 \\
Bilateral filters (1) & Filter size = 5 × 5, $\sigma_c$ = 0.1, $\sigma_s$ = 5 \\
Bilateral filters (2) & Filter size = 5 × 5, $\sigma_c$ = 1.0, $\sigma_s$ = 5 \\
Median blur filter & Filter size = 5 × 5 \\
Gaussian blur filter (1) & Filter size = 5 × 5, $\sigma$ = 0.5 \\
Gaussian blur filter (2) & Filter size = 5 × 5, $\sigma$ = 1.5 \\
\hline\hline
\end{tabular}}
  \caption{Information on each classical filter, where $\sigma_c$, $\sigma_s$, and $\sigma$ represent the filter standard deviation in the color space, coordinate space, and Gaussian kernel, respectively.}
  \label{tab:Table3}
  \end{center}
\end{table}

\subsection{Evaluation 1: Noisy bluff body flow data with Gaussian noise at various noise levels}
\label{Evaluation 1: Noisy bluff body flow data with Gaussian noise at various noise levels} 

In this subsection, noisy bluff body flow data with Gaussian noises at 1/SNR = 0.1, 0.5, and 1.5 were used to train the DAE, where each noisy data set includes 1200 snapshots. In the testing process, noisy bluff body flow data at various noise levels from the training data were used to check the denoising performance and DAE generalization. In addition to the test using training data, noisy data at 1/SNR = 0.3 and 1 were used for the interpolation test, whose noise levels were within the noise level range of the training data. Noisy data at 1/SNR = 0.05, 0.075, 1.75, and 2 were used for the extrapolation test, whose noise levels were out of the noise level range of the training data. \par

\begin{figure}
\centering 
\includegraphics[angle=0, trim=0 0 0 0, width=1\textwidth]{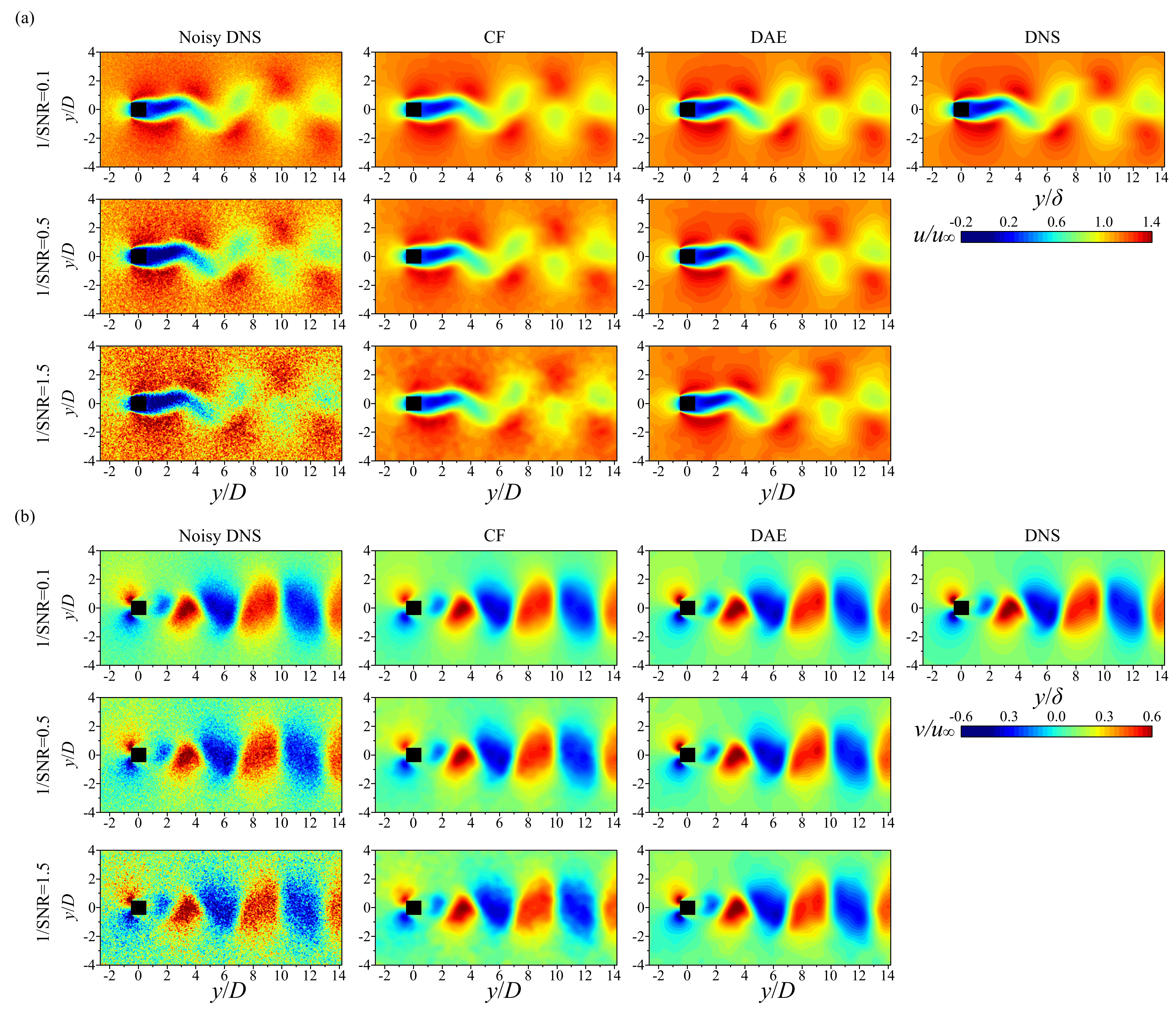}
\caption[]{Denoised instantaneous contours of bluff body flow with Gaussian noise at various noise levels: (a) streamwise velocity and (b) spanwise velocity.}
\label{fig:6}
\end{figure}

\begin{figure}
\centering 
\includegraphics[angle=0, trim=0 0 0 0, width=1\textwidth]{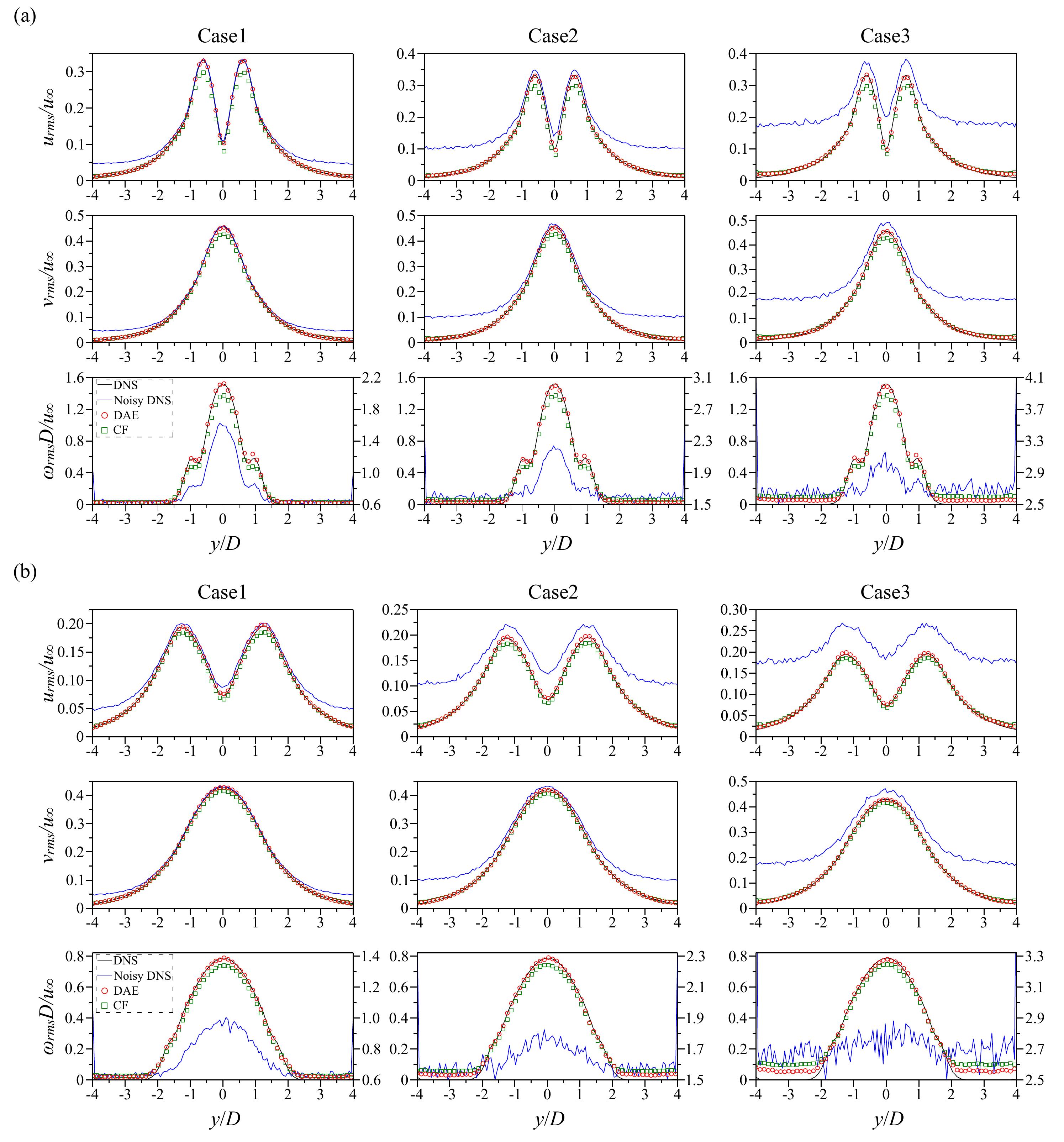}
\caption[]{Root-mean-square profiles of streamwise velocity, spanwise velocity, and vorticity of bluff body flow: (a) data sampling at $x$/$d$ = 1, and (b) data sampling at $x$/$d$ = 6. Cases 1, 2, and 3 are denoising cases with Gaussian noise at 1/SNR = 0.1, 0.5, and 1.5, respectively. For the vorticity plots, the right vertical axis represents noisy DNS.}
\label{fig:7}
\end{figure}

\begin{figure}
\centering 
\includegraphics[angle=0, trim=0 0 0 0, width=1\textwidth]{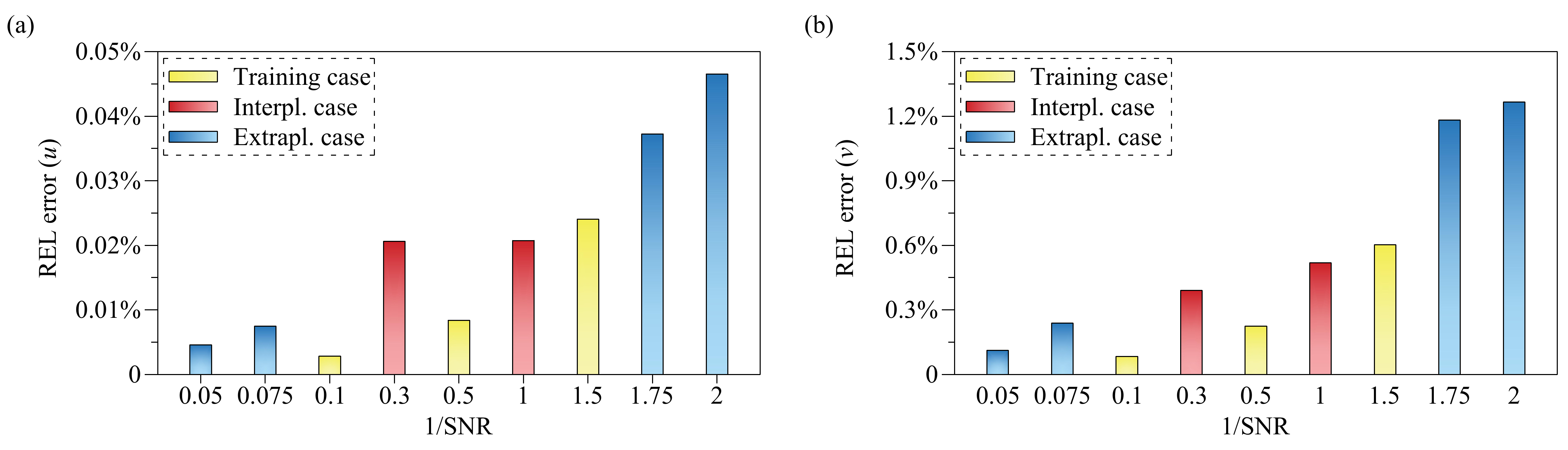}
\caption[]{Relative error of training, interpolation, and extrapolation cases: (a) and (b) are the relative errors of streamwise and spanwise velocity, respectively.}
\label{fig:8}
\end{figure}

Figure \ref{fig:6} displays the denoised instantaneous contours of bluff body flow with Gaussian noise at 1/SNR = 0.1, 0.5, and 1.5 (training case). The DAE outperformed CF in denoising, especially in the high noise level case. DAE allowed the noise to be removed. The flow fields could be accurately reconstructed. Figure \ref{fig:7} displays the root-mean-square profiles of velocity fields sampling at various positions of $x$ axis. The results obtained from DAE were consistent with those of DNS. However, the results of the CF exhibit apparent deviations compared with DNS. Figure \ref{fig:8} displays the relative error of training, interpolation, and extrapolation cases. The relative errors revealed an increasing trend with the increase in the noise level. However, all the errors were minor, where the maximum error of $u$ is 0.05\% and 1.5\% for $v$. Although the interpolation case never attends DAE training, the errors of the interpolation case were similar to those of the training case at 1/SNR =1.5. Notably, the extrapolation case not only absented in DAE training but also featured noise levels beyond the range encountered in the training cases. However, the extrapolation case could be effectively denoised, resulting in a low error level. Therefore, Figure. 8 indicates that the proposed denoising methods exhibit excellent generalization that is suitable for denoising noisy data at a range of noise levels, even if the noise level is out of the range of the training case.\par

\subsection{Evaluation 2: Noisy bluff body flow data with noises of three types}
\label{Evaluation 2: Noisy bluff body flow data with noises of three types} 

In this subsection, noisy bluff body flow data with various noises, that is, Gaussian noise, salt-and-pepper noise, and speckle noise at the same noise level (1/SNR = 0.5), were passed to train the DAE, where each noisy data includes 1200 snapshots. \par

\begin{figure}
\centering 
\includegraphics[angle=0, trim=0 0 0 0, width=1\textwidth]{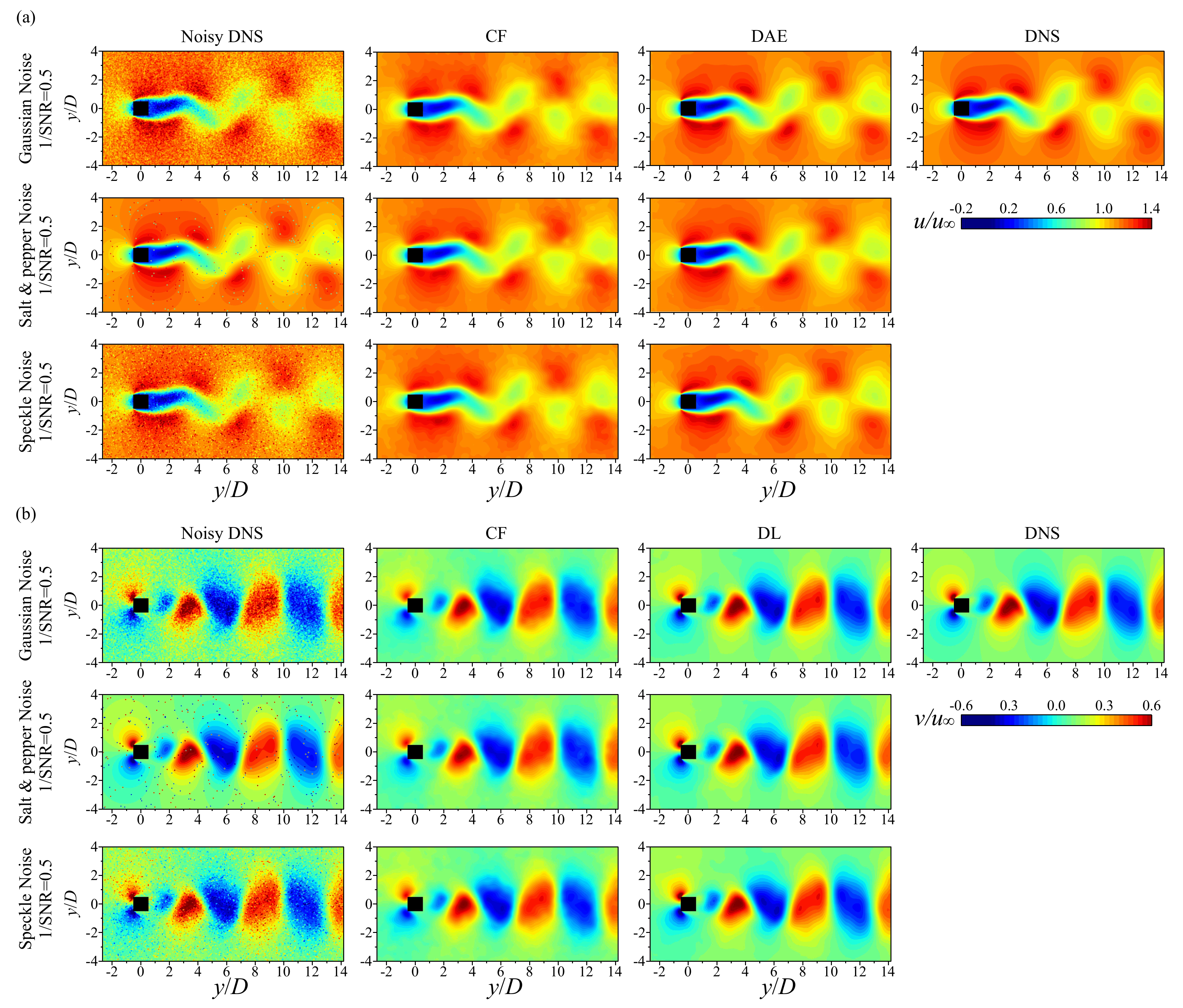}
\caption[]{Denoised instantaneous contours of the bluff body flow with three noises at noise level 1/SNR = 0.5: (a) streamwise velocity and (b) spanwise velocity.}
\label{fig:9}
\end{figure}

\begin{figure}
\centering 
\includegraphics[angle=0, trim=0 0 0 0, width=1\textwidth]{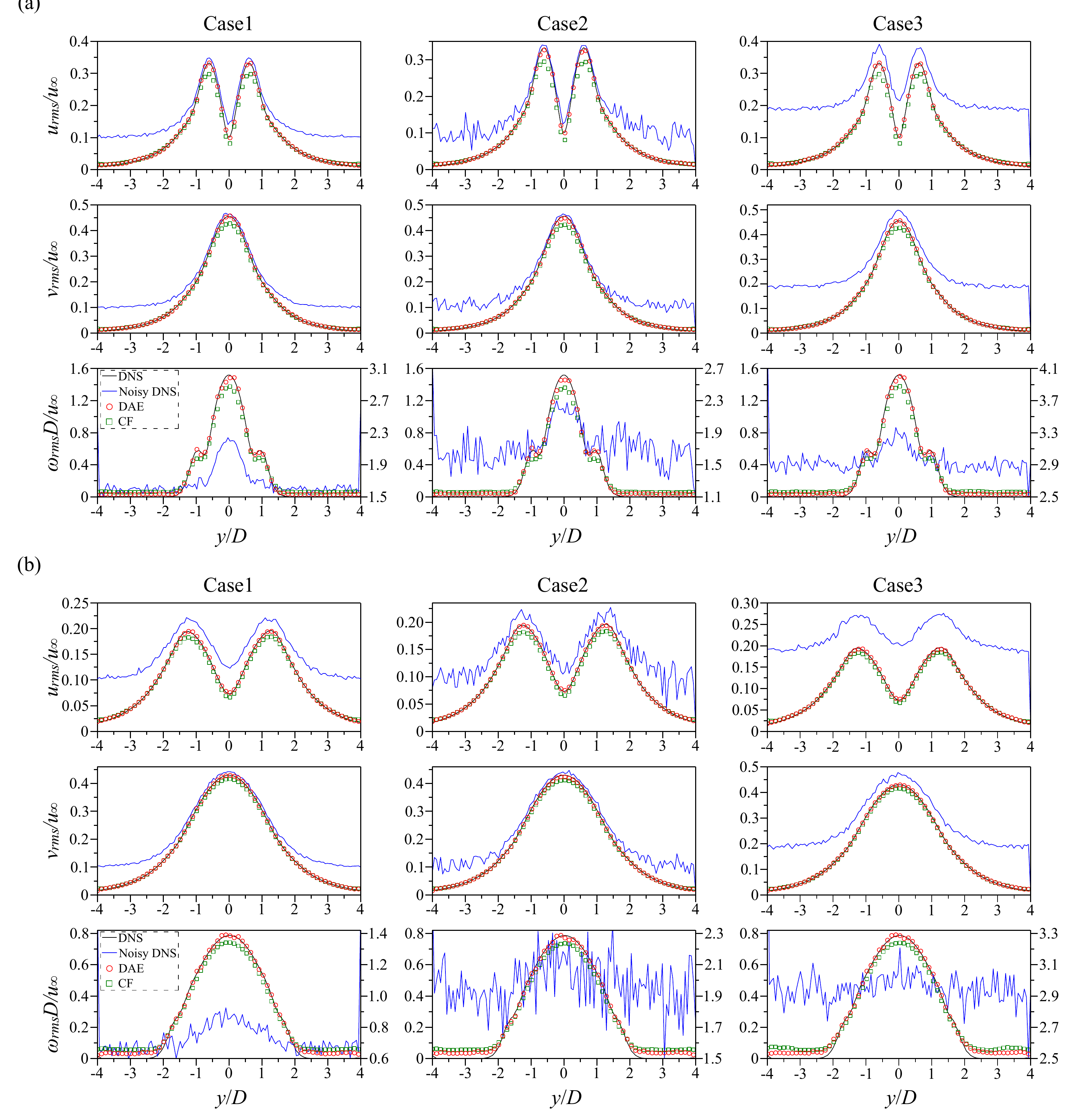}
\caption[]{Root-mean-square profiles of the streamwise velocity, spanwise velocity, and vorticity of the bluff body flow: (a) data sampling at $x$/$d$ = 1, and (b) data sampling at $x$/$d$ = 6. Cases 1, 2, and 3 are denoising cases with Gaussian noise, salt-and-pepper noise, and speckle noise, respectively. For vorticity, the right vertical axis represents the noisy DNS.}
\label{fig:10}
\end{figure}

Figure \ref{fig:9} displays the denoise instantaneous contours of the bluff body flow, contaminated with three types of noise at a noise level of 1/SNR = 0.5. The results denoised by DAE appear smoother than those processed by CF. Moreover, although the CF struggled with denoising salt-and-pepper and speckle noise, DAE consistently removed these three distinct noise types. Furthermore, as depicted in Figure \ref{fig:10}, the statistical outcomes derived from the DL method exhibit higher accuracy than those obtained through CF.\par

\subsection{Evaluation 3: Noisy channel flow data at three Reynolds numbers}
\label{Evaluation 3: Noisy channel flow data at three Reynolds numbers} 

\begin{figure}
\centering 
\includegraphics[angle=0, trim=0 0 0 0, width=1\textwidth]{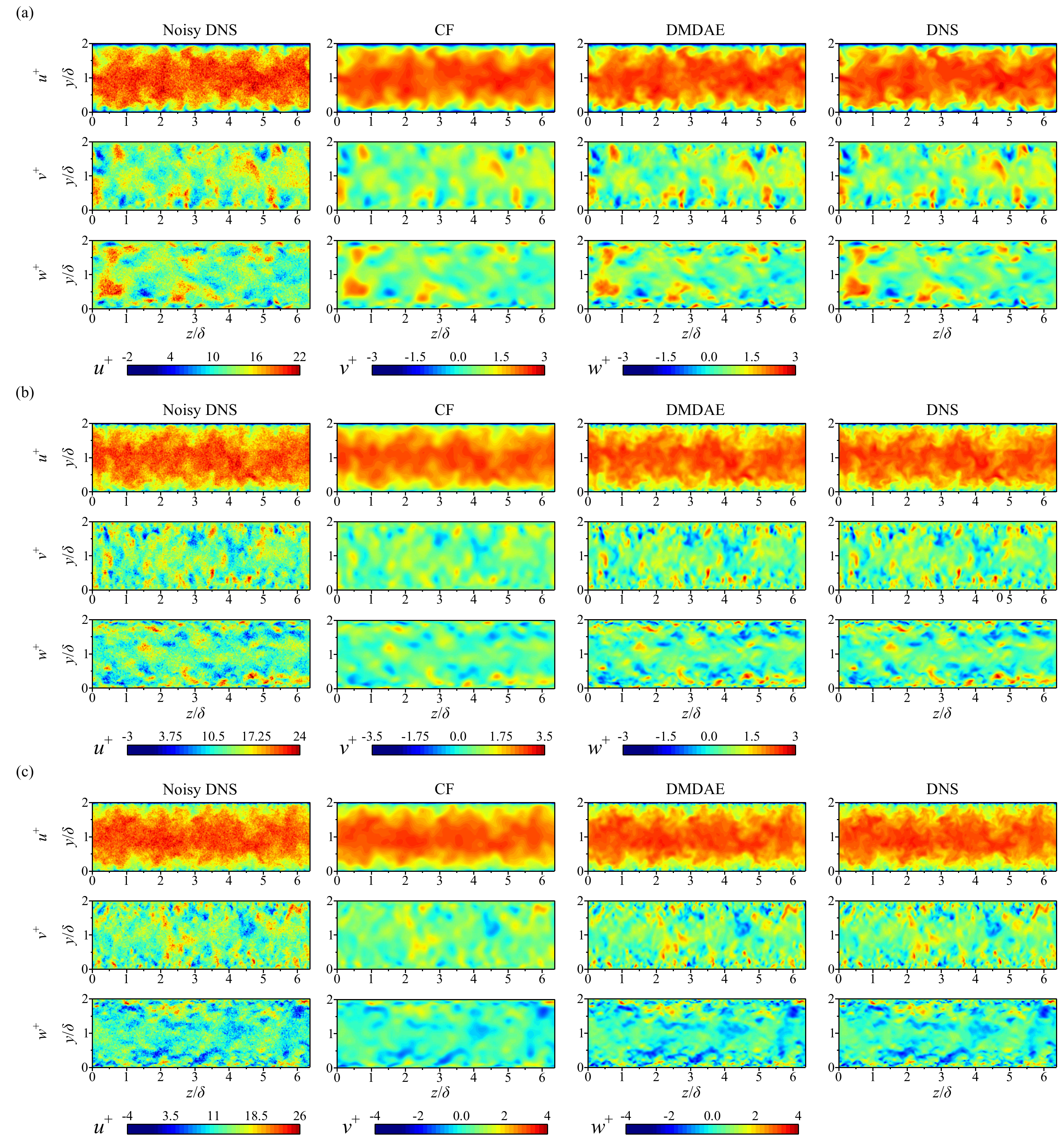}
\caption[]{Denoised instantaneous contours of the channel flow with various Reynolds numbers at noise level 1/SNR = 0.5: (a) streamwise velocity, (b) spanwise velocity, and (c) wall-normal velocity.}
\label{fig:11}
\end{figure}

\begin{figure}
\centering 
\includegraphics[angle=0, trim=0 0 0 0, width=1\textwidth]{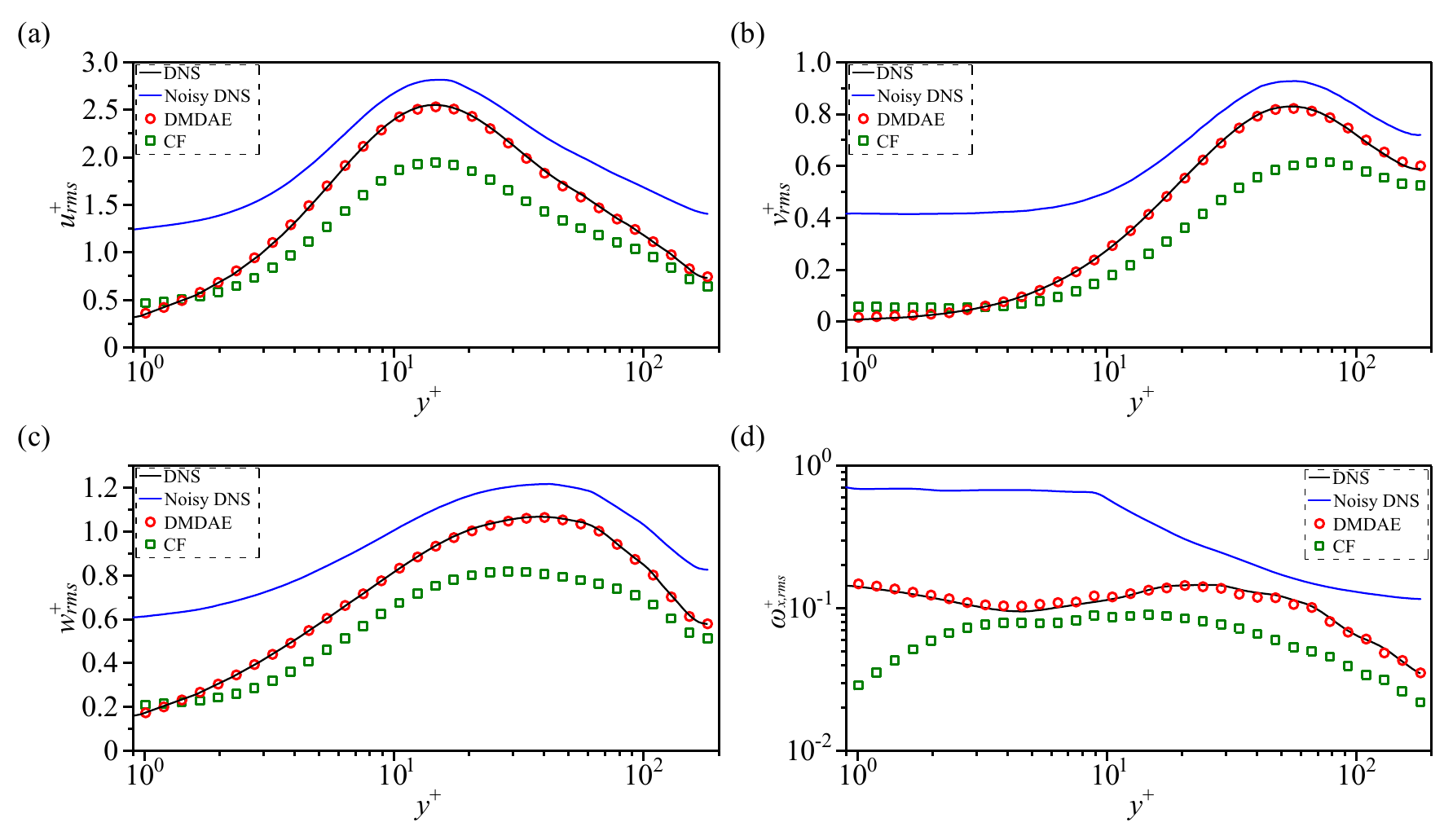}
\caption[]{Root-mean-square profiles of (a) streamwise velocity, (b) spanwise velocity, (c) wall-normal velocity, (d) and streamwise vorticity of the channel flow at $Re_\tau$ = 180.}
\label{fig:12}
\end{figure}

\begin{figure}
\centering 
\includegraphics[angle=0, trim=0 0 0 0, width=1\textwidth]{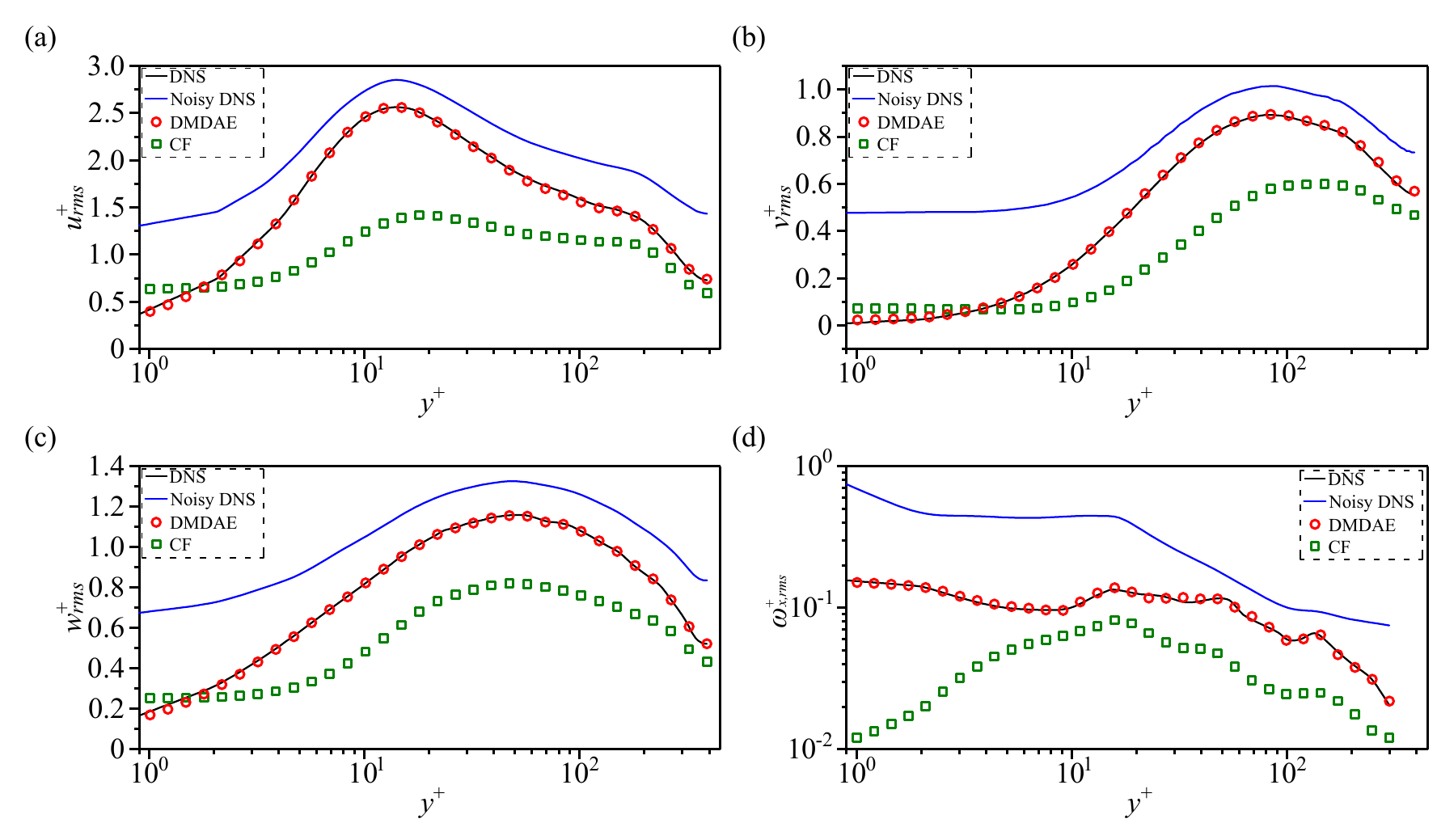}
\caption[]{Root-mean-square profiles of (a) streamwise velocity, (b) spanwise velocity, (c) wall-normal velocity, (d) and streamwise vorticity of the channel flow at $Re_\tau$ = 395.}
\label{fig:13}
\end{figure}

\begin{figure}
\centering 
\includegraphics[angle=0, trim=0 0 0 0, width=1\textwidth]{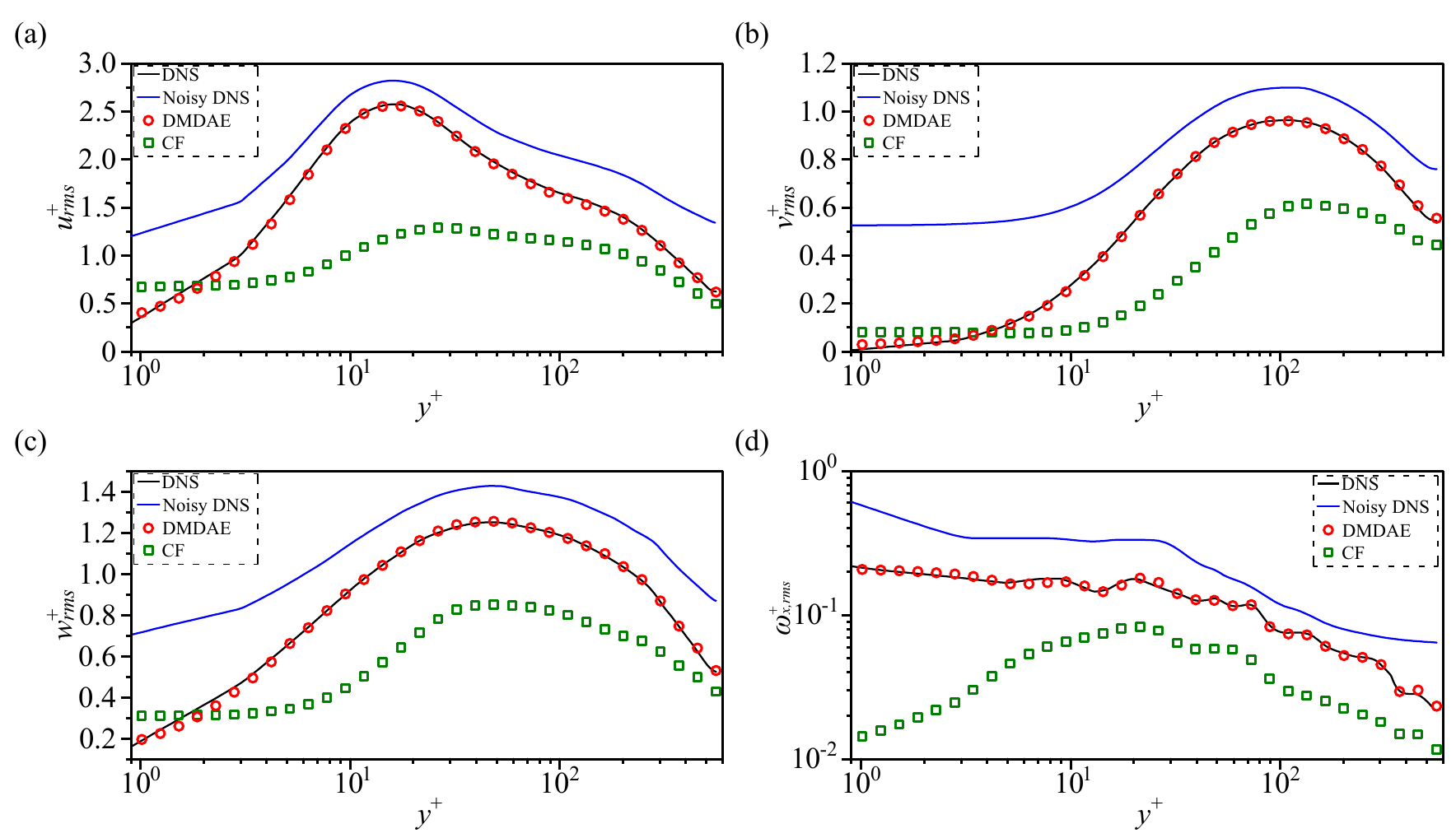}
\caption[]{Root-mean-square profiles of (a) streamwise velocity, (b) spanwise velocity, (c) wall-normal velocity, (d) and streamwise vorticity of the channel flow at $Re_\tau$ = 550.}
\label{fig:14}
\end{figure}

\begin{figure}
\centering 
\includegraphics[angle=0, trim=0 0 0 0, width=1\textwidth]{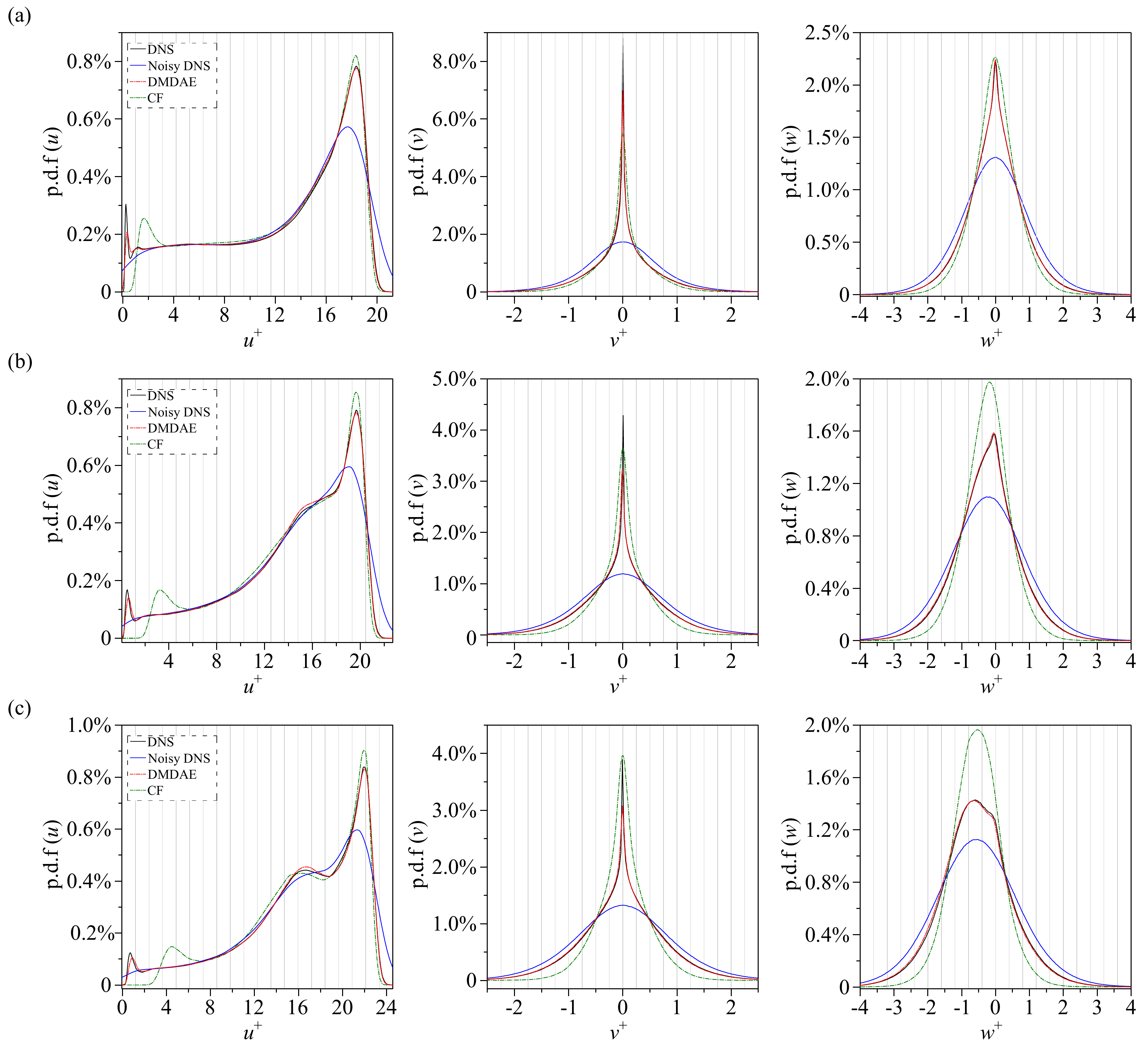}
\caption[]{Probability density function of the velocity components for the channel flow at (a) $Re_\tau$ = 180, (b) $Re_\tau$ = 395, and (c) $Re_\tau$ = 550.}
\label{fig:15}
\end{figure}

\begin{figure}
\centering 
\includegraphics[angle=0, trim=0 0 0 0, width=1\textwidth]{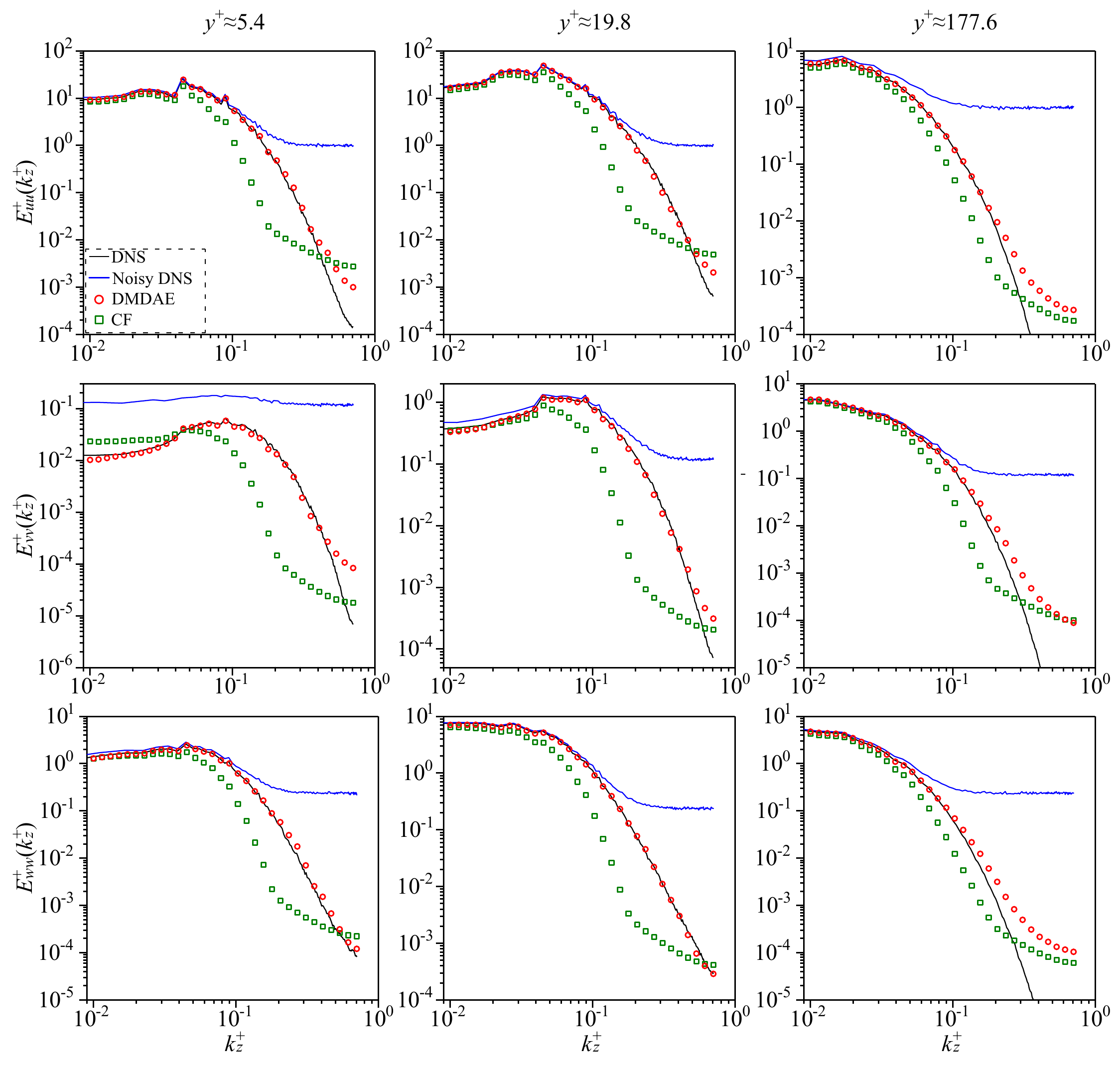}
\caption[]{Spanwise energy spectra of the velocity components for the turbulent channel flow at (a) $Re_\tau$ = 180.}
\label{fig:16}
\end{figure}

\begin{figure}
\centering 
\includegraphics[angle=0, trim=0 0 0 0, width=1\textwidth]{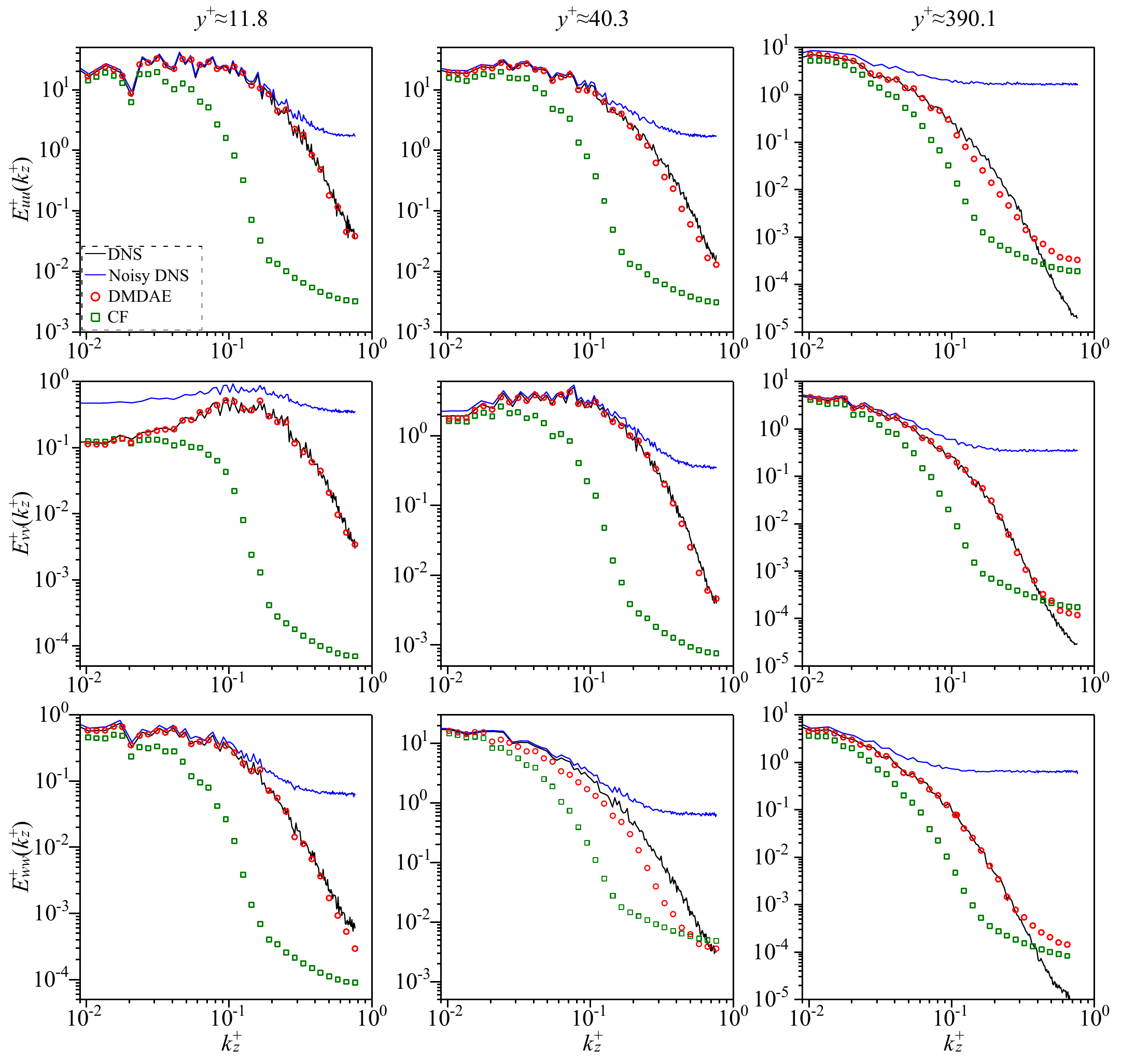}
\caption[]{Spanwise energy spectra of the velocity components for the turbulent channel flow at (a) $Re_\tau$ = 395.}
\label{fig:17}
\end{figure}

\begin{figure}
\centering 
\includegraphics[angle=0, trim=0 0 0 0, width=1\textwidth]{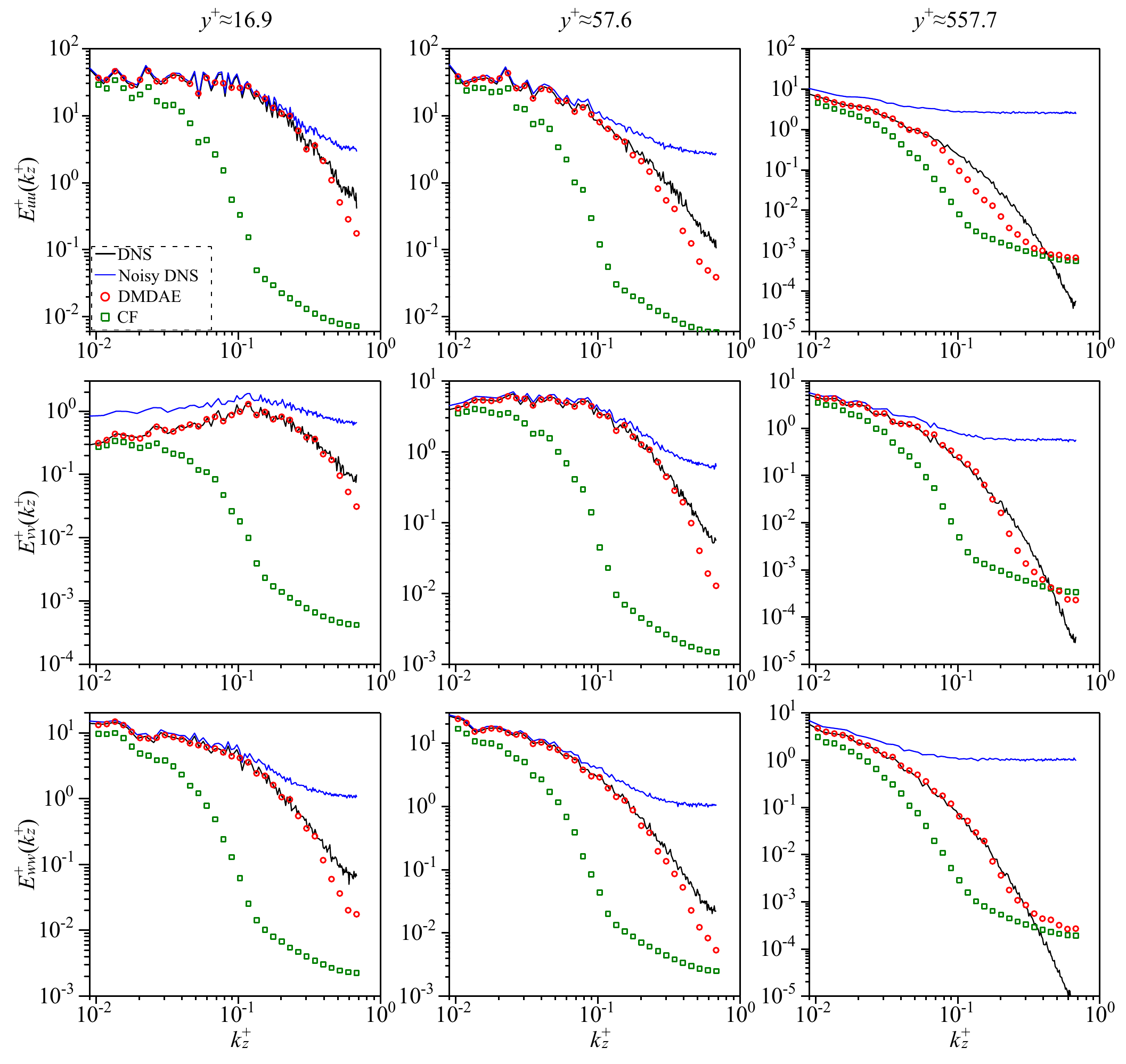}
\caption[]{Spanwise energy spectra of the velocity components for the turbulent channel flow at (a) $Re_\tau$ = 550.}
\label{fig:18}
\end{figure}

In this subsection, DMDAE was used to denoise turbulent channel flows at $Re_\tau$ = 180, 395, and 550, which presents a higher level of complexity than the laminar flow cases. The dataset used for training DMDAE comprises cases of noisy channel flow data with Gaussian noise at 1/SNR = 0.5, each containing 1200 snapshots. Various outcomes, including instantaneous contours and turbulent statistics, are illustrated to demonstrate that the denoise data numerically and physically align with DNS data.
The instantaneous contours (Figure \ref{fig:11}) indicate that DMDAE effectively denoises and reconstructs the three-channel flow cases. By contrast, the CF method could not accurately distinguish flow features from noise, erroneously eliminating flow features as noise. Figures \ref{fig:12}, \ref{fig:13}, and \ref{fig:14} demonstrate the velocity and vorticity RMS profiles for the three-channel flow cases. The outcomes obtained from DMDAE were consistent with the DNS results, whereas the results from CF exhibited deviations compared with DNS. Figure \ref{fig:15} displays the probability density function ($p.d.f$) of the velocity components for channel flow. The velocity distribution curves of the DL results were consistent with the ground truth. Furthermore, the most pronounced deviation in the CF results occurred in the case of $Re_\tau$ = 550, where the flow was more chaotic and challenging to denoise. Figures\ref{fig:16}, \ref{fig:17}, and \ref{fig:18} reveal the spanwise energy spectra of the velocity components. Energy spectra results revealed that conventional denoising filters perform poorly on turbulent flow denoising as the eddies of channel flow were mistakenly eliminated. The proposed DL method performs better in ensuring that the denoise results are consistent with the DNS data.

\subsection{Evaluation 4: noisy PIV data}
\label{Evaluation 4: noisy PIV data}

\begin{figure}
\centering 
\includegraphics[angle=0, trim=0 0 0 0, width=1\textwidth]{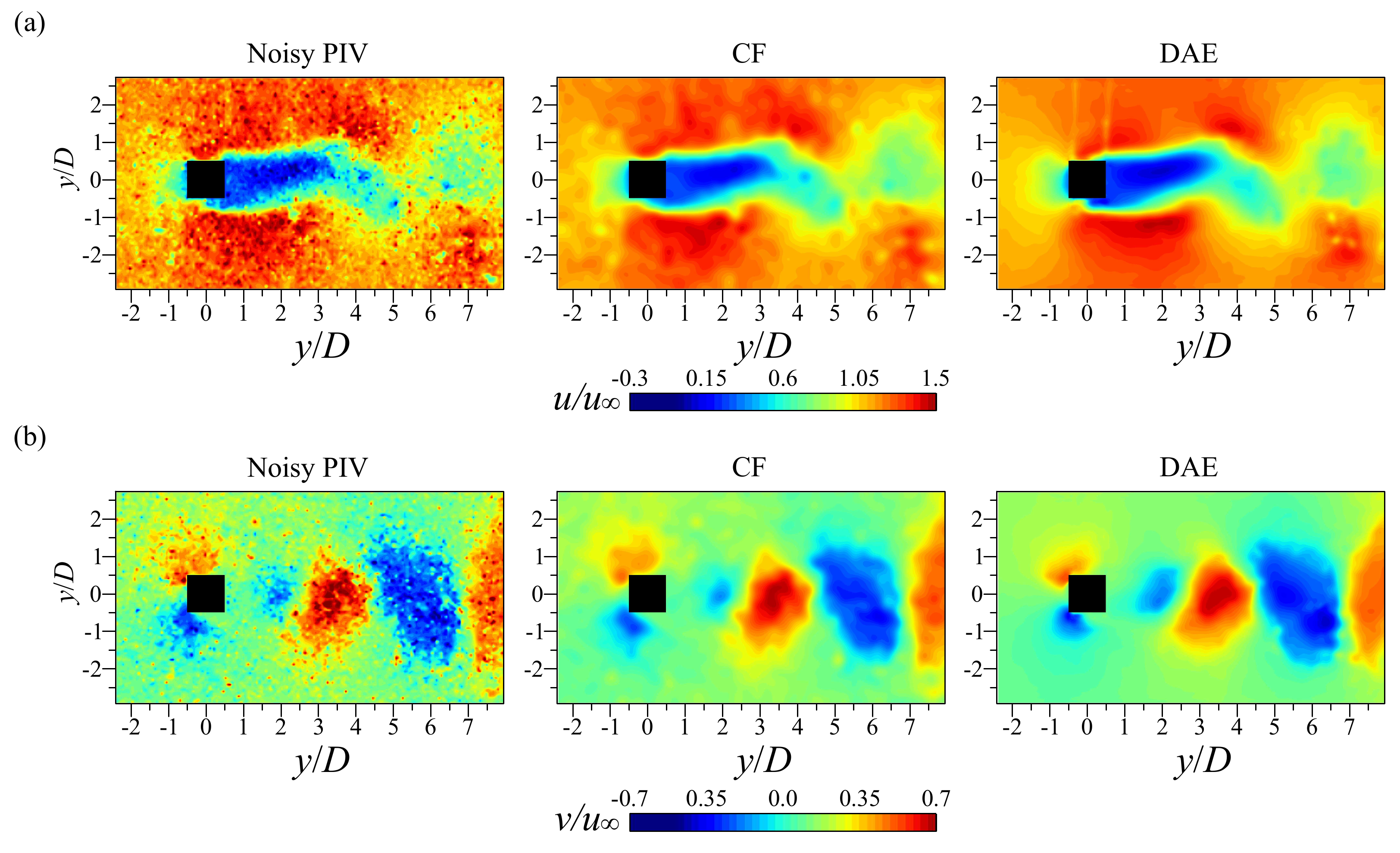}
\caption[]{Denoised instantaneous contours of noisy PIV: (a) streamwise velocity and (b) spanwise velocity.}
\label{fig:19}
\end{figure}

\begin{figure}
\centering 
\includegraphics[angle=0, trim=0 0 0 0, width=1\textwidth]{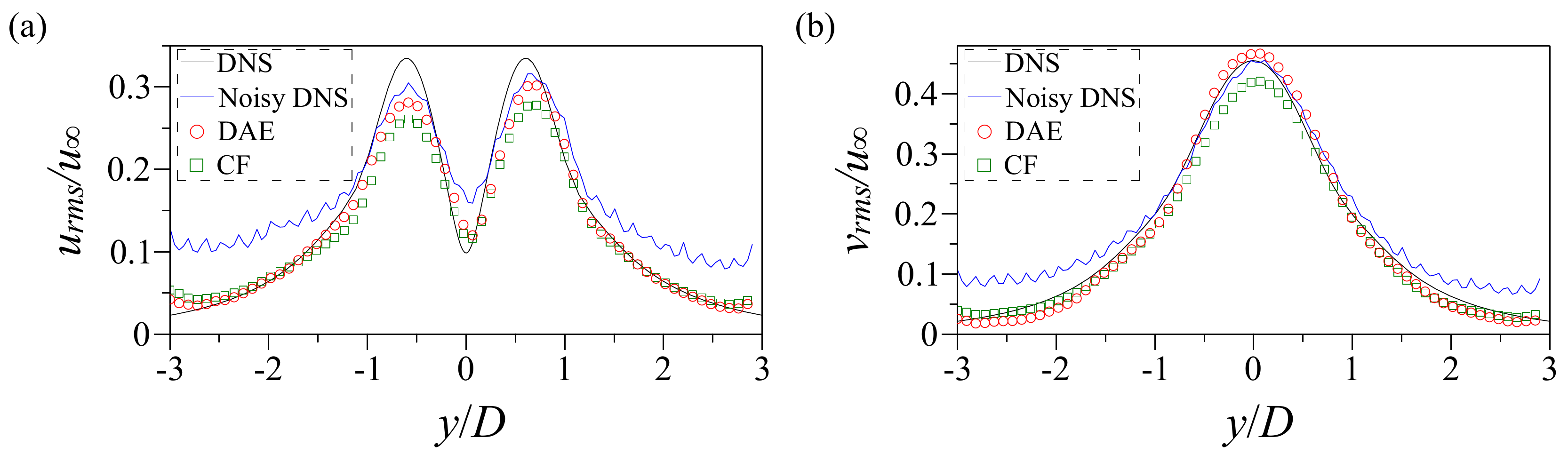}
\caption[]{Root-mean-square profiles of streamwise velocity (a) and spanwise velocity (b), where data sampling was performed at $x$/$d$ = 1.}
\label{fig:20}
\end{figure}

In this subsection, the denoising performance of the DAE was examined using noisy PIV data. Figure \ref{fig:19} displays the denoise instantaneous contours of the PIV data. Comparing the contours of denoise and DNS data reveals that the proposed DL model can remove most noise from noisy PIV data. However, the CF cannot denoise sufficiently where some large-scale noise exists in the flow field. By contrast, DAE cannot remove the shadow region near the bluff body. The shadow region exists in every snapshot of the PIV data; therefore, DAE regards the shadow as a part of the flow and retains it. Thus, DAE can only remove the noise randomly appearing but not restore the permanent corruption within the flow field. Figure \ref{fig:20} represents the RMS profiles of the velocity components, where a DNS case having the same $Re_d$ as the PIV data is also plotted. The figure reveals that the denoised data are closer to the DNS results after denoising using the DL model.

\section{Conclusions}
\label{Conclusion}
This study proposed two efficient flow field denoising DL models, namely DAE and DMDAE. The models were developed based on the self-learning principle. Therefore, the training process did not require clean data as the label. The bottleneck-shaped autoencoder with a small latent space could denoise noisy flow data. The convolutional and max-pooling layers in the encoder played a funnel role in filtering out the noise. Benefiting from eliminating noise in the encoder, the decoder would not reconstruct the noise but instead output denoised flow fields.
Furthermore, DNS bluff body flow data and DNS turbulent channel flow data at various Reynolds numbers with synthetic noise were applied to evaluate the denoising capability of the proposed DL models. Three noise types, Gaussian, salt-and-pepper, and speckle noise, were used in this study. Furthermore, PIV bluff body data with natural noise were used to examine DAE denoising performance. For comparison, some classical noise filters denoise the flow fields. Four evaluations were conducted in this study.
First, noisy bluff body flow data with Gaussian noises at 1/SNR = 0.1, 0.5, and 1.5 were used to train the model. In addition to regular testing, interpolation testing (noise level within the range of 0.1 to 1.5) and extrapolation (noise level out of 0.1 to 1.5) were adopted. The results revealed that DAE achieved excellent denoising capability and generalization. All the noise data cases, even the noise level not in the training data range, were well denoised. Second, noisy bluff body flow data with three noises were simultaneously used to train DAE. DAE deleted all the types of noises. Next, turbulent channel flow data with Gaussian noise at 1/SNR = 0.5 were adapted to evaluate the denoising performance for the turbulent flow by using DMDAE. The noisy channel flow data at three Reynolds numbers ($Re_\tau$ = 180, 395, and 550) were used to train the DMDAE. The instantaneous contours and statistical results indicated that DMDAE could effectively denoise noisy turbulent flow. Finally, the last evaluation proved that DAE could also denoise noisy PIV data with actual noise. Furthermore, the proposed DL method outperformed the classical denoising filter in denoising. In particular, classical denoising filters performed worst in turbulent channel flow cases, where the CF could not distinguish the small eddies and flow features from noise.
The proposed DAE and DMDAE could be efficient and applicable in denoising in the fluid dynamic field, especially from their power denoising capability and generalization.


\begin{acknowledgments}
This work was supported by 'Human Resources Program in Energy Technology' of the Korea Institute of Energy Technology Evaluation and Planning (KETEP), granted financial resource from the Ministry of Trade, Industry \& Energy, Republic of Korea (no. 20214000000140). This work was supported by the Korean Cancer Research Institute grant (2024).
\end{acknowledgments}

\section*{Data Availability}
The data that supports the findings of this study are available within this article.

\nocite{}
\bibliography{my-bib}
\end{document}